\documentclass[onecolumn,prf,
 amsmath,amssymb,
]{revtex4-2}

\DeclareMathOperator{\csch}{csch}
\usepackage{hyperref}
\usepackage{xcolor}

\usepackage{graphicx}
\usepackage{dcolumn}
\usepackage{bm}

\usepackage{amssymb}
\usepackage{natbib}

\begin{document}

\preprint{APS/123-QED}

\title{Revisiting Crossflow-Based Stabilization in Channel Flows} 

\author{Muhammad Abdullah}
 \email{muhabd@seas.upenn.edu}
\author{George I. Park}%
\altaffiliation{Corresponding author}
 \email{gipark@seas.upenn.edu}
\affiliation{%
 Department of Mechanical Engineering \& Applied Mechanics, University of Pennsylvania.
}%

\date{\today}

\begin{abstract}
Stabilization schemes in wall-bounded flows often invoke fluid transpiration through porous boundaries. While these have been extensively validated for external flows, their efficacy in channels, particularly from the standpoint of non-modal perturbations, is yet to be demonstrated. Here, we show that crossflow strengths previously considered ``ideal'' for optimizing stability in channels in fact admit strong non-modal energy amplification. We begin by supplementing existing modal calculations and then show via the resolvent that extremely strong and potentially unfeasible crossflows are required to suppress non-modal growth in linearly stable regimes. Investigation of unforced algebraic growth paints a similar picture. Here, a component-wise budget analysis reveals that energy redistribution through pressure-velocity correlations plays an important role in driving energy growth/decay. The superposition of a moving wall is also considered, and it is shown that while energy amplification generally worsens, it can potentially be suppressed beyond critical strengths of the crossflow. However, these flow regimes are marred by rapidly declining mass transport, rendering their ultimate utility questionable. Our results suggest that crossflow-based stabilization might not be useful in internal flows.
\end{abstract}

\maketitle

\section{\label{sec:introduction} Introduction}

The control of fluid flows in channels is among the most complex open problems in classical theory, a precise understanding of which is yet to be established despite multiple decades of research. The shift from a benign laminar state to the complex chaos of turbulence is, to all appearances, highly non-linear, and incurs key structural and mechanistic changes in the processes driving mass and momentum transport. These modifications, depending on the exact setting, can be both desirable or undesirable, so that strategies to suppress -- or even instigate -- them are typically of significant interest. In this work, our aim is to reassess the efficacy of one such scheme: the superposition of a homogeneous, vertical (i.e., in the wall-normal direction) crossflow, such as that generated by a steady, spatially uniform transpiration of fluid through a porous boundary (or boundaries). Such systems have widespread utility, for example, in filtration processes, medical apparatus, and various geophysical and astrophysical phenomena, although, surprisingly, it is applications in external flows that are more frequently cited in the literature (e.g. modulating flow separation or delaying inflectional instabilities over swept wings; see \cite{joslin1998aircraft} or \cite{KRISHNAN201724}). 

To preface with a broader perspective, the debate on transition and instability in fluid flows has undoubtedly been a long-standing one. Here, the standard analysis proceeds by deriving an eigenvalue problem from the equations of motion linearized around some base state of interest, usually laminar. Eigenmodes with positive growth rates suffer exponential amplification in time, and a critical value, $Re_c$, of the Reynolds number defines the threshold below which such an instability cannot survive. For simple parabolic profiles, such as those encountered in rectilinear pressure-driven flows, this approach predicts $Re_c\approx 5772$, with the Reynolds number defined based on the channel half-height $h$ and the center-line velocity $U_p$. As expected, the inclusion of crossflow, described by its own Reynolds number, say $R_v$, dramatically alters the picture, and the first few treatments may perhaps be attributed to \cite{hains} and \cite{sheppard}. Both authors resolved, at least for small to moderate $R_v$, a significant increase in $Re_c$ (again based on $U_p$), suggesting that the overarching influence of a net throughflow was very much stabilizing. Evidently, this remained the prevailing opinion until \cite{fransson_and_alfredsson} proposed an alternative non-dimensionalization of the dynamical variables, setting the ``actual'' streamwise maximum as opposed to $U_p$ as the reference velocity unit. Because the injected wall-normal flux dampens the base streamwise component, this framework, they argued, preserved the distinction between the base velocity magnitude and the base velocity distribution when investigating stability. With this scaling, \cite{fransson_and_alfredsson} demonstrated pockets of stabilization and destabilization, called ``branches'' in their terminology. To recount their example of choice, the wavenumber $\alpha = 1$ at $Re = 6000$, known to exhibit a positive growth rate for Poiseuille flow, becomes increasingly stable until a turning point of $R_v\approx 3.4$. Thereafter, the trend reverses and the instability is restored for $R_v\gtrapprox42.5$, only to be suppressed once again for $R_v \approx \mathcal{O}(10^2)$.

Apart from the initial study by \cite{hains}, the work of \cite{GUHA_2010} appears to be the first to provide a rigorous extension to the case of moving (upper) boundaries, the so-called Couette-Poiseuille flows, which present interesting stability characteristics in and of themselves. In particular, adopting the dimensionless ratio $\xi = U_c/U_p$ (the equivalent proxy in \cite{GUHA_2010} is $k$), where $U_c$ is the wall speed, \cite{potter_1966} showed that complete linear stability -- $Re_c\to\infty$ -- could be achieved for $\xi\gtrapprox0.7$, termed the ``cutoff''. This is, of course, not surprising since the Couette flow itself is known to be linearly stable to infinitesimal perturbations for various rheological models; see, for example, \cite{Romanov1973}. At any rate, in the presence of crossflow, \cite{GUHA_2010} reported trends largely similar to the $\xi=0$ case, except that the cutoff threshold could now be reduced to, say, $\xi \approx 0.2$ using intermediate $R_v$ (a claim that, as they noted, was only valid for $Re\leq 10^6$, the upper limit of their numerical study). As special instances of this base flow, ``generalized" Couette-like profiles have also been investigated by \cite{nicoudandangilella}; these are precisely linear and develop when the influence of the crossflow in the streamwise momentum budget is neutralized by a suitably chosen pressure gradient. In the presence of viscosity, these profiles exhibit complete linear stability up to $R_v\approx 24$, beyond which the critical Reynolds number can descend to as low as the relatively mild $Re_c\approx725$ \citep{nicoudandangilella}. Even in the inviscid
(Rayleigh) limit, excellently treated by \cite{Kerswell_2015}, a perturbation has the potential to grow. Finally, we remark that the spatially developing (i.e., in the streamwise direction) analog has also been of interest and has been rigorously explored using Berman-type similarity solutions in the works of \cite{Zaturska_1988} and \cite{Watson_Banks_Zaturska_Drazin_1990}.

Consequently, crossflow-laden channel flows demonstrate rich stability portraits and form an interesting testbed for studying transition and disturbance amplification. From a purely application-oriented point of view, previous work, particularly on spatially non-developing cases, can be summarized in the statement by \cite{GUHA_2010}, who posited that modest wall velocities $\xi$ coupled with weak-to-moderate crossflows strike the optimal balance between stability and practicality. We argue here -- and this is our main contribution -- that this is not necessarily the case. In particular, we cite the non-normality of the linearized operator, typical of inertia-dominated flows, as the main culprit. Since the unstable eigenvalues are only really relevant at sufficiently large time horizons, when all other modal contributions have more or less decayed, one can rarely accurately judge the stability of a system from its spectrum alone. In shear flows, non-modal growth mechanisms are generally more dominant and therefore more informative, which is a well-established result in hydrodynamic stability theory \citep{tref_pseudospec, TrefethenEmbree, schmidstability, brandt_lift_up}. Surprisingly, despite their apparent simplicity, a comprehensive analysis on non-modal perturbations in (internal) crossflow-laden flows is yet to be performed, at least to the best of our knowledge. One relevant work in this regard seems to be that of \cite{Chen_2021}, who investigated transient, unforced, energy amplification using the base flow of \cite{fransson_and_alfredsson}. However, the scope of their study was rather narrow, and, as we discuss below, leaves much room for further investigation. On the other hand, the single-plate, zero-pressure-gradient, analog -- the so-called asymptotic suction boundary layer (ASBL) -- has received much more attention \citep{FRANSSON2003259, asbl_experiment, CHERUBINI2015246}, and we make comparisons where appropriate.

We structure this paper as follows.\,\,Section \ref{sec:problem_formulation} introduces the base flow and describes our analysis frameworks. Section \ref{sec:modal_analysis} augments previous discussion on eigenvalues, while Section \ref{sec:nonmodal_analysis} presents a systematic summary of our non-modal calculations, with a particular focus on the purely pressure-driven case. Section \ref{sec:wall_velocity} investigates the effects of wall motion, while Section \ref{sec:conclusion} offers conclusions.

\section{\label{sec:problem_formulation} Problem Formulation}
\subsection{\label{ssec:base_velocity_profiles} The Base Flow}

\begin{figure}
\centering
\includegraphics[width=\textwidth]{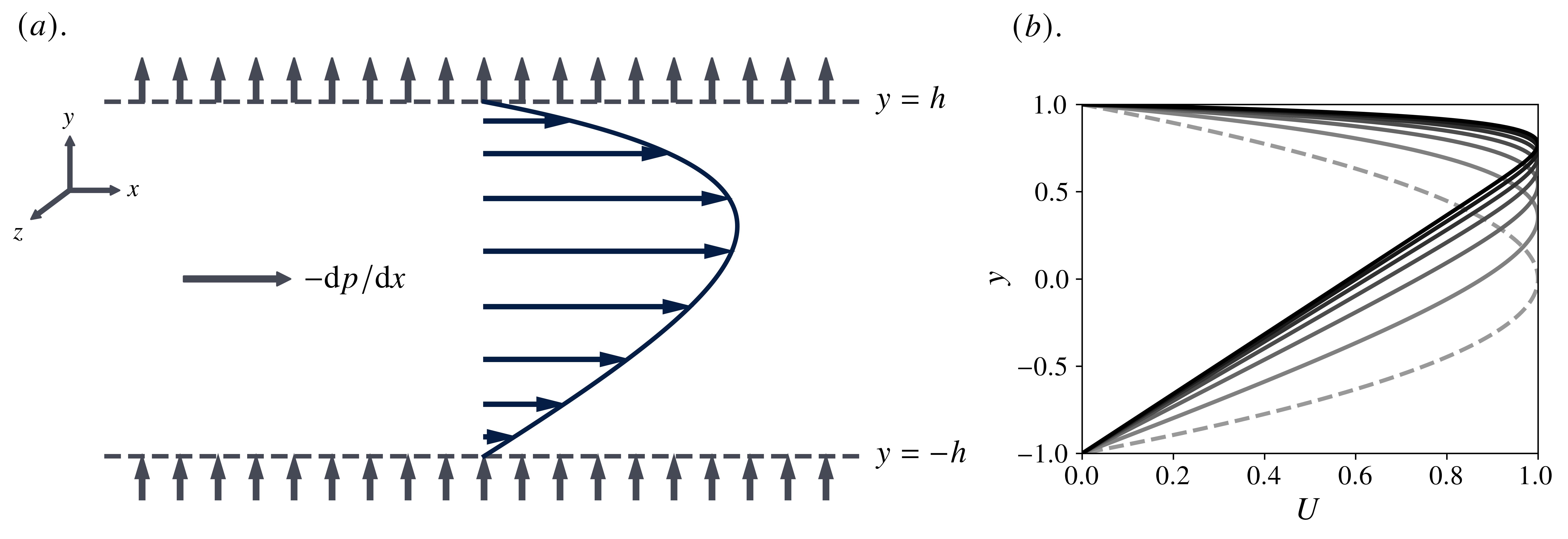}
\caption{$(a)$, a diagram of the present flow geometry; $(b)$, streamwise velocity profiles for $R_v = 0$ (dashed line) to $R_v = 15$ in increments of 2.5. Lighter to darker shades represent increasing $R_v$. Note that, contrary to the convention adopted by \cite{fransson_and_alfredsson}, the direction of the crossflow here is from bottom to top.}
\label{fig:geometry}
\end{figure}

We begin our analysis with the standard Navier-Stokes equations. In dimensional terms, these read
\begin{align}
\label{eqn:nse_dim}
    \dfrac{\mathrm{D}\boldsymbol{u}}{\mathrm{D}t} & = \dfrac{1}{\rho}\left(\nabla\cdot\mathsf{T}\right), \\
\label{eqn:cont}
\nabla\cdot\boldsymbol{u} & = 0.
\end{align}
where $\mathrm{D}/\mathrm{D}t\equiv \partial/\partial t + \left(\boldsymbol{u}\cdot\nabla\right)$ is the total derivative, $\mathsf{T} = -p\mathsf{I} + 2\mu\mathsf{e}$ is the Cauchy stress, and $\mathsf{e} = \left(\nabla\boldsymbol{u} + \left(\nabla\boldsymbol{u}\right)^\intercal\right)/2$ is the symmetric rate-of-strain tensor.\,\,Here, Equation (\ref{eqn:cont}) encodes the usual incompressibility constraint. The system of interest in this study can be summarized in the schematic presented in Figure \ref{fig:geometry}$(a)$. An incompressible Newtonian fluid, confined between two rigid, porous, infinite boundaries positioned at $y=\pm h$, is driven by a constant streamwise pressure gradient $-\mathrm{d}p/\mathrm{d}x > 0$. A vertical crossflow is imposed by introducing a uniform injection, $V_i$, and suction, $V_s$, of fluid through the lower and upper walls, respectively. We assume a fully-developed flow, so that $\partial/\partial x = \partial/\partial z\equiv 0$ and the compatibility of the boundary conditions with the continuity equation demands $V_i = V_s = V_0$.

As noted earlier, the sensible selection of a reference velocity for this system is rather nuanced. The usual (and admittedly immediate) choice is $U_p$, the center-line velocity for the Poiseuille flow in the absence of crossflow. However, since the streamwise velocity is directly coupled to -- and decreases -- with the crossflow strength, for sufficiently large $V_0$, $U_p$ is not a physically meaningful metric. Therefore, following \cite{fransson_and_alfredsson}, we adopt $U_m$, the streamwise maximum, as our characteristic velocity. With the channel half-width $h$ as our length scale, the associated Reynolds number becomes $Re = U_mh/\nu$, where $\nu$ is the kinematic viscosity, and the dimensionless form of the governing equations can be written as
\begin{align}
\label{eqn:nse_nondim}
\boldsymbol{u}_t + \left(\boldsymbol{u}\cdot\nabla\right)\boldsymbol{u} & = -\nabla p + \dfrac{1}{Re}\nabla^2\boldsymbol{u}, \\
\nabla\cdot\boldsymbol{u} & = 0.
\end{align}
Furthermore, with the crossflow Reynolds number $R_v = V_0 h/\nu$, the stationary laminar profile becomes
\begin{equation}
\label{eqn:base_state}
    \boldsymbol{U} = \begin{pmatrix}
        U\left(y, R_v\right) & V_0 / U_m & 0
    \end{pmatrix}^\intercal = \begin{pmatrix}
        U\left(y, R_v\right) & R_v/Re & 0
    \end{pmatrix}^\intercal,
\end{equation}
where
\begin{equation}
\label{eqn:streamwise_profile}
    U\left(y, R_v\right) = \frac{R_v(y-\csch R_v e^{R_v y}+\coth R_v)}{R_v\coth R_v - 1 - \log\left(R_v\csch R_v\right)}.
\end{equation}
Figure \ref{fig:geometry}$(b)$ illustrates $U$ plotted against the non-dimensional $y$-coordinate for some representative values of $R_v$. Since the (non-dimensional) streamwise maximum has been fixed to unity, the primary effect of a stronger crossflow is to skew the profile in the positive wall-normal direction. Consequently, as noted by \cite{fransson_and_alfredsson}, a very thin boundary layer develops near the upper (suction) wall, leaving an approximately linear profile throughout the remainder of the channel. In fact, one can show that
\begin{equation}
    \lim_{R_v\to\infty} U\left(y, R_v\right) = \dfrac{1}{2}\left(1+y\right),
\end{equation}
in the channel bulk, which, save for the different top-wall boundary condition, is precisely the well-known Couette profile for viscous flow between two parallel surfaces in relative motion. Thus, as $R_v\to\infty$, complete linear stability may be expected. On the other hand, the absence of crossflow, $R_v = 0$, yields the Poiseuille flow, although establishing this directly from Equation (\ref{eqn:streamwise_profile}) requires additional care, since $U\left(y, R_v\right)$ is not defined in this limit. In fact, $R_v = 0$ is a removable singularity for $U$, and the appropriate power expansion
\begin{equation}
    U\left(y, R_v\right) = (1-y^2) + \dfrac{R_v}{3}(y-y^3) + O(R_v^2),
\end{equation}
informs the smooth continuation of $U\left(y, R_v\right)$ over all $R_v\in\mathbb{R}$. 

\subsection{\label{ssec:linearized_equations} The Linearized Equations}
To formulate the stability problem, we rewrite the Navier-Stokes equations in operator format
\begin{equation}
    \dfrac{\partial \bm{\chi}}{\partial t} = g\left(\bm{\chi}\right)
\end{equation}
where $\bm{\chi}$ is the state vector and $g$ is a non-linear differential operator. Representing the base profile as $\overline{\bm{\chi}}$, we consider a set of infinitesimal fluctuations $\bm{\chi}^\prime$ and perform a Jacobian linearization of $g$ around $\overline{\bm{\chi}}$. The result is a system of linearized evolution equations for $\bm{\chi}^\prime$, which, in the traditional fourth-order formulation involving perturbations around the base state of the wall-normal velocity/vorticity  
$\left(v^\prime, \eta^\prime\right)$, reads
\begin{align}
\label{eqn:os}
\left[\left(\dfrac{\partial}{\partial t} + U\dfrac{\partial}{\partial x} + V\dfrac{\partial}{\partial y}\right)\nabla^2 - \dfrac{\mathrm{d}^2 U}{\mathrm{d}y^2}\dfrac{\partial}{\partial x}- \dfrac{1}{Re}\nabla^4\right]v' & = 0,\\
\label{eqn:sq}
\left[\dfrac{\partial}{\partial t} + U\dfrac{\partial}{\partial x} + V\dfrac{\partial}{\partial y} - \dfrac{1} {Re}\nabla^2\right]\eta'+\dfrac{\mathrm{d}U}{\mathrm{d}y}\dfrac{\partial v'}{\partial z}&= 0.
\end{align}
Here, $\nabla^4\left\langle\cdot\right\rangle = \nabla^2\left(\nabla^2\left\langle\cdot\right\rangle\right)$ is the standard bi-harmonic operator acting in Cartesian space, and $U$ and $V$ are the streamwise and wall-normal components of the background flow, Equation (\ref{eqn:base_state}). The appropriate boundary conditions follow from the non-slip and impermeability restrictions at the wall
\begin{equation}
    \left.v^\prime\right|_{y= \pm 1}  = \left.\eta^\prime\right|_{y= \pm 1}  = \left.\partial v^\prime/\partial y\right|_{y= \pm 1} = 0.
\end{equation}
In what follows, the prime notation is dropped with the understanding that all discussion is based on the disturbance field, unless explicitly noted. We exploit the spatial homogeneity in the wall-parallel directions by applying a wave-like ansatz
\begin{equation}
\label{eqn:normal_modes}
    \left(v, \eta\right) = \left(\widetilde{v}\left(y, t\right), \widetilde{\eta}\left(y, t\right)\right)e^{i\left(\alpha x+\beta z\right)},
\end{equation}
where $\alpha,\beta\in\mathbb{R}$ represent the spatial wavenumbers. The result is the Orr-Sommerfeld-Squire initial-value problem
\begin{equation}
\label{eqn:ivp}
\mathsf{L}\boldsymbol{q} = -\dfrac{\partial}{\partial t}\mathsf{M}\boldsymbol{q} \implies \dfrac{\partial \boldsymbol{q}}{\partial t} = (-\mathsf{M}^{-1}\mathsf{L})\boldsymbol{q} \implies \dfrac{\partial \boldsymbol{q}}{\partial t} = \mathsf{S}^\prime\boldsymbol{q}.
\end{equation}
In Equation (\ref{eqn:ivp}), $\boldsymbol{q} = \begin{pmatrix} \widetilde{v} & \widetilde{\eta} \end{pmatrix}^\intercal$ and, by denoting $\mathsf{D}\equiv \partial/\partial y$ and $k^2 = \alpha^2+\beta^2$, the block operators $\mathsf{L}$ and $\mathsf{M}$ are
\begin{equation}\label{eqn:orr_somm}
    \mathsf{L} = \begin{pmatrix}
        \mathsf{L}_\mathrm{OS} & 0 \\ i\beta \mathsf{D}U  & \mathsf{L}_\mathrm{SQ}
    \end{pmatrix},\quad \mathsf{M} = \begin{pmatrix}
        \mathsf{D}^2 - k^2 & 0 \\ 0 & 1
    \end{pmatrix},
\end{equation}
where
\begin{align}
\label{eqn:os_operator}
    \mathsf{L}_\mathrm{OS} & = \left(i\alpha U+V\mathsf{D}\right)(\mathsf{D}^2-k^2) - i\alpha \mathsf{D}^2 U-\dfrac{1}{Re}(\mathsf{D}^2 - k^2)^2, \\
    \label{eqn:sq_operator}
    \mathsf{L}_\mathrm{SQ} & = i\alpha U +V\mathsf{D} -\dfrac{1}{ Re}(\mathsf{D}^2-k^2).
\end{align}
The condition for Hurwitz stability can then be written as $\mathfrak{R}\left(\lambda\right) < 0$ for all $\lambda\in\Lambda\left(\mathsf{S}^\prime\right)$, where $\Lambda\left(\mathsf{S}^\prime\right)$ denotes the spectrum of $\mathsf{S}^\prime$. Typically, these eigenvalues are subsequently related to a set of complex circular frequencies $\omega = \omega_r + i\omega_i$, such that $\lambda = -i\omega$, rephrasing the stability constraint as $\omega_i < 0$ \citep{schmid_annrev}. This completes the definition of the normal mode in Equation (\ref{eqn:normal_modes}), that is,
\begin{equation}
    \left(v, \eta\right) = \left(\widetilde{v}\left(y, t\right), \widetilde{\eta}\left(y, t\right)\right)e^{i\left(\alpha x+\beta z\right)} = \left(\hat{v}\left(y\right), \hat{\eta}\left(y\right)\right)e^{i\left(\alpha x+\beta z-\omega t\right)}.
\end{equation}
In this approach, we are particularly interested in the manifold of neutral stability, defined here by the locus
\begin{equation}
    \omega_i\left(\alpha, \beta, Re, R_v\right) = 0.
\end{equation}
It is well-known, however, that a naive analysis of the spectral abscissa in this way is often very limiting, especially in shear flows, because it provides predictions of flow stability only at asymptotically long times. In contrast, significant energy growth can be initiated by non-modal mechanisms operating on much shorter time scales, potentially violating the linear assumption prior to the emergence of the unstable eigenmode, if any, and encouraging the onset of non-linear interactions (possibly even turbulence). This amplification usually supersedes the often weak growth rates associated with modal solutions and therefore evades identification in a treatment based solely on eigenvalues \citep{tref_pseudospec, schmid_annrev}. In fact, even in the simplest examples of shear flows, the transition to turbulence is highly sub-critical, that is, it occurs well below any threshold for the Reynolds number as predicted by eigenvalue theory. A model for this behavior lies in the non-normality of $\mathsf{S}^\prime$, whose commutator with its adjoint ${\mathsf{S}^\prime}^\dagger$ need not vanish 
\begin{equation}
   [\mathsf{S}^\prime, {\mathsf{S}^\prime}^\dagger] = \mathsf{S}^\prime {\mathsf{S}^\prime}^\dagger -  {\mathsf{S}^\prime}^\dagger\mathsf{S}^\prime\neq \mathsf{0}.
\end{equation}
As a corollary, $\mathsf{S}^\prime$ admits oblique (non-orthogonal) eigenfunctions that, in a basis expansion, can allow for finite-time energy growth, evidently in both sub- and super-critical parameter regimes. This (purely linear) mechanism is often identified as a likely motivator for the so-called \textit{bypass} route to turbulence \citep{butler_3d_opt_pert, brandt_lift_up}, and has even found utility in elucidating key physical mechanisms driving fully turbulent flows; see, for example, \cite{cossu1, cossu2, cherubini1}.

To explore the potential for non-modal growth here, we work from the most general case, that is, when the initial-value problem in Equation (\ref{eqn:ivp}) is driven by a time-harmonic forcing $\mathsf{F}\left(y, t\right) = \mathsf{f}\left(y\right)e^{-i\upsilon t}$. Thus, we write
\begin{equation}
    \dfrac{\partial \boldsymbol{q}}{\partial t} = -i\mathsf{S}\boldsymbol{q} + \mathsf{F} \implies \dfrac{\partial \boldsymbol{q}}{\partial t} = -i\mathsf{S}\boldsymbol{q} + \mathsf{f}\left(y\right)e^{-i\upsilon t},
\end{equation}
where $\upsilon\in\mathbb{C}$ and $\mathsf{S} = i\mathsf{S}^{\prime}$. Since $\mathsf{S}^{\prime}$ is time-independent, the system response is
\begin{equation}
        \boldsymbol{q}\left(t\right) = \mathsf{\Phi}\left(t, 0\right)\boldsymbol{q}\left(0\right) - ie^{-i\upsilon t}\left(\mathsf{S} - \upsilon\mathsf{I}\right)^{-1}\mathsf{f}\left(y\right),
\end{equation}
where $\mathsf{\Phi}\left(t, 0\right)\equiv e^{-i\mathsf{S}t}$ is the solution propagator, which maps the initial state of the system $\boldsymbol{q}_0$ at $t^{\prime} = 0$ to its value at $t^{\prime} = t$, and $\mathsf{R} \equiv \left(\mathsf{S} - \upsilon\mathsf{I}\right)^{-1}$ is the resolvent of $\mathsf{S}$. Consider now the special case where $\mathsf{F} = \mathsf{0}$; under appropriate norms in the input and output spaces, the gain can be defined as
\begin{equation}
    G\left(\alpha, \beta, Re, R_v, t\right) = \sup_{\boldsymbol{q}_0\neq 0}\dfrac{\left\lVert \boldsymbol{q}\right\rVert_{\mathrm{out}}^2}{\left\lVert \boldsymbol{q}_0\right\rVert_{\mathrm{in}}^2},
\end{equation}
and represents, at time $t$, the largest energy growth optimized over all possible initial conditions having unit norm \citep{schmidstability}. As a physically meaningful metric, the energy norm is adopted here (see, for example, \cite{butler_3d_opt_pert}), so that $\left\lVert \cdot\right\rVert_{\mathrm{in}} = \left\lVert \cdot\right\rVert_{\mathrm{out}} = \left\lVert \cdot\right\rVert_E$ and
\begin{equation}
\label{eqn:energy_norm}
    \left\lVert \boldsymbol{q}\right\rVert_E^2 = \int_{-1}^1\widetilde{v}^\dagger\widetilde{v} + \dfrac{1}{k^2}\left(\widetilde{\eta}^\dagger\widetilde{\eta} + \mathsf{D}\widetilde{v}^\dagger\mathsf{D}\widetilde{v}\right)\,\mathrm{d}y,
\end{equation}
where we have restricted attention to the real component of the disturbance. Therefore, it follows that $G = \left\lVert \mathsf{\Phi}\left(t,0\right)\right\rVert_E^2$. Hereafter, there are two equally valid approaches: (i), convert directly to a weighted 2-norm through a similarity transformation incorporating information on the non-uniform grid spacing \citep{TrefethenEmbree} or (ii), project onto the space of eigenfunctions using the Gramian matrix $\mathsf{M} = \mathsf{W}^\dagger\mathsf{W} \succ 0$ \citep{schmidstability}. At any rate, only a standard singular value decomposition is required. In this setting, the right and left singular functions represent, respectively, the initial condition and response pair that achieve the gain $G$ at time $t$.

On the other hand, if $\mathsf{F}\neq \mathsf{0}$, then assuming asymptotic stability of $\mathsf{S}^\prime$, the long-time response reduces to
\begin{equation}
    \boldsymbol{q}\left(t\right) = -ie^{-i\upsilon t}\left(\mathsf{S} - \upsilon\mathsf{I}\right)^{-1}\mathsf{f}\left(y\right)
\end{equation}
Here, the resolvent $\mathsf{R}$ becomes the quantity of interest, encoding important information about the spectral properties of the system. In particular, if the excitation frequency is resonant, so that $\upsilon\in\Lambda\left(\mathsf{S}\right)$, the resolvent is ill-defined and its norm $\left\lVert \mathsf{R}\right\rVert_E$ tends to infinity. However, for systems whose dynamics are governed by non-normal operators, this norm may be large even when $\upsilon$ is merely pseudoresonant, that is, $\upsilon\notin\Lambda\left(\mathsf{S}\right)$. If $\upsilon$ is restricted to real frequencies, the analysis can, in a sense, be physically motivated, with the resolvent describing the perturbed linear operator resulting from, for example, exogeneous vibrations or experimental imperfections \citep{tref_pseudospec}. Generalizing to the complex plane allows for the definition of the so-called $\epsilon$-pseudospectrum, the set of values given by
\begin{equation}
    \Lambda_\epsilon\left(\mathsf{S}\right) = \left\{\upsilon\in\mathbb{C} \colon \left\lVert \mathsf{R}\right\rVert_E\geq 1/\epsilon\right\}
\end{equation}
For normal operators, the $\epsilon$-pseudospectra, at least under an appropriate 2-norm as chosen here, correspond to closed $\epsilon$-balls centered around the spectrum. Non-normality, on the other hand, allows for more complicated pseudospectral boundaries, and the extent to which they protrude into the stable half of the complex plane can have important implications for energy growth in the unforced initial-value problem -- see, for example, \cite{reddyschmidhenn} or \cite{TrefethenEmbree}.

To discretize the stability operators, we implement a standard Chebyshev pseudospectral method written in \texttt{Python}.\,\,Our in-house solver has been extensively validated against classical rectilinear geometries, including the (Newtonian and Oldroyd-B) Poiseuille flow, the Couette flow, and the Couette-Poiseuille flow, and has recently been used in the linear analysis of three-dimensional boundary layers \citep{abdullah2024linear}. In most of the calculations presented here, unless specifically noted, we employed $N = 256$ Chebyshev modes, resulting in a $\left(2N+2\right)\times\left(2N+2\right)$ matrix system. We determined that this resolution was sufficient -- and occasionally necessary -- to achieve convergence for the values of $R_v$ treated here. All modal and non-modal calculations were performed in \texttt{SciPy}. To accelerate our output, we additionally scaled to an embarrassingly parallel workload using the \texttt{Python} module \texttt{Ray} \citep{ray_paper}. Finally, the $\epsilon$-pseudospectra were created using Eigentools \citep{eigentools}.

\section{\label{sec:modal_analysis} Modal Analysis}

We begin by supplementing classical perspectives on the modal stability of this flow. We define the critical Reynolds number, $Re_c$, as the smallest value of the Reynolds number below which the flow is linearly stable. At this threshold, a disturbance, described by the critical wavenumber $\left(\alpha_c,\beta_c\right)$ must achieve neutral stability. Despite a non-zero mean velocity component in the wall-normal direction, Squire's transformation \citep{squires} remains applicable, allowing us to set $\beta_c = 0$ \textit{a priori} and restrict attention to transversal modes. In contrast to the purely streamwise ($V = 0$) case, however, the Squire operator (see Equations (\ref{eqn:orr_somm}) and (\ref{eqn:sq_operator})) cannot be ignored, since the associated eigenmodes need not be damped. To confirm this, we follow \cite{schmidstability} by converting to a formulation involving the complex phase speed, $c = \omega / \alpha = c_r + ic_i$, and multiplying the homogeneous Squire problem by $\eta$. Integrating across $y$, leveraging the associated Dirichlet conditions, and isolating imaginary components, we find
\begin{equation}
    \label{eqn:squire_modes}
    c_i\int_{-1}^1\left|\hat{\eta}\right|^2\mathrm{d}y = -\mathfrak{R}\left[\int_{-1}^1\hat{\eta}^\dagger\mathsf{D}\hat{\eta}\,\mathrm{d}y\right]\dfrac{V}{\alpha} - \dfrac{1}{\alpha Re}\int_{-1}^1\left|\mathsf{D}\hat{\eta}\right|^2 + k^2\left|\hat{\eta}\right|^2\,\mathrm{d}y
\end{equation}
Assuming $\alpha > 0$, the second term on the right-hand side of Equation (\ref{eqn:squire_modes}) is necessarily negative, although the first may or may not be. The latter, of course, vanishes when $V = 0$, ensuring $c_i < 0$. Thus, for our purposes, we consider the complete Orr-Sommerfeld-Squire system when investigating modal stability.
\begin{figure}
\centering
\includegraphics[width=0.95\textwidth]{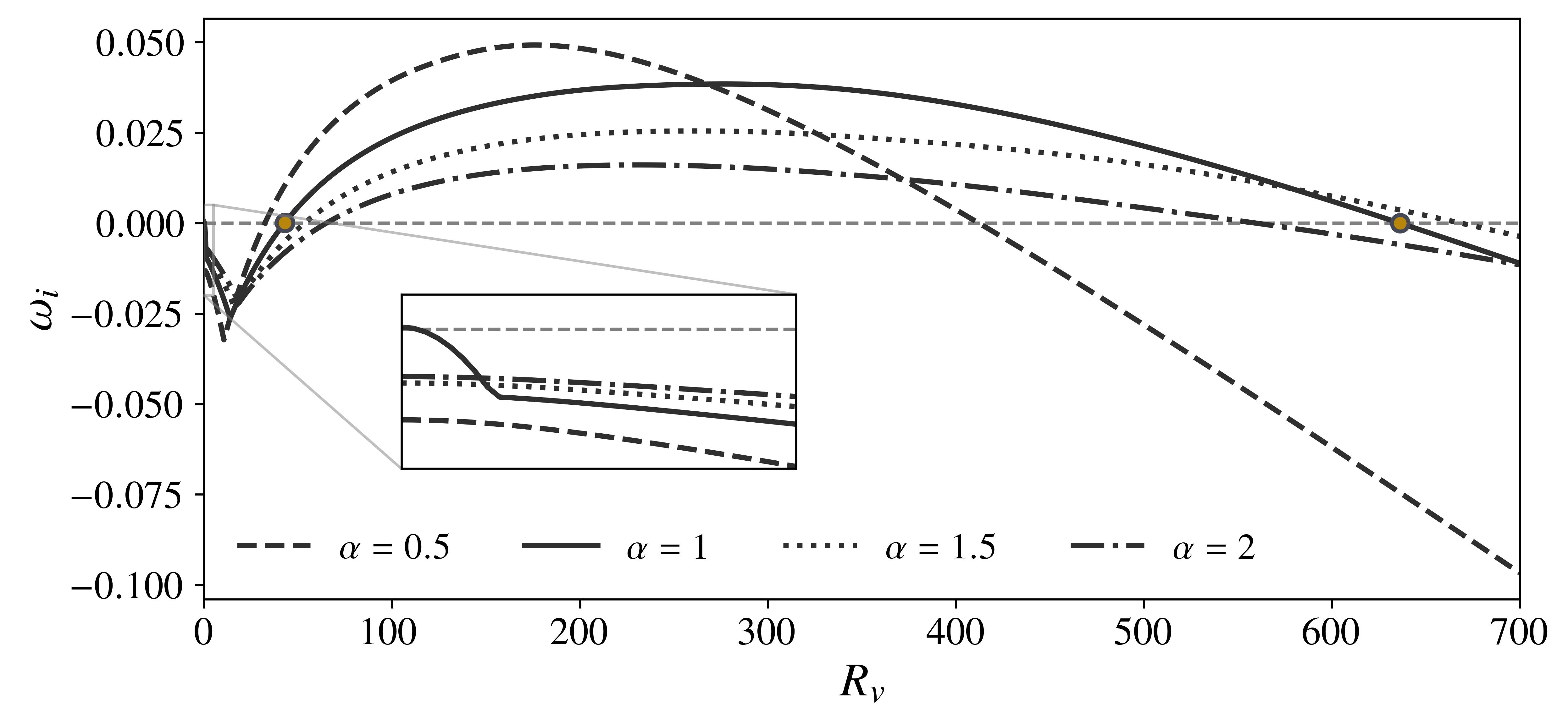}
\caption{At $Re = 6000$, the growth rates $\omega_i$ versus $R_v$ for the most unstable eigenmode corresponding to $\alpha \in \left\{0.5, 1, 1.5, 2\right\}$. The gray dashed line denotes the stability boundary $\omega_i = 0$. For $\alpha=1$, the Branch-I and Branch-II $R_v$ have been emphasized through circles.}
\label{fig:modal_growth_rates}
\end{figure}

To preface the ensuing discussion, we cite a key finding from \cite{fransson_and_alfredsson}, referring in the process to Figure \ref{fig:modal_growth_rates}, which plots at $Re = 6000$ the growth rate $\omega_i$ corresponding to the least stable eigenmode for various representative choices of the streamwise wavenumber $\alpha$. \cite{fransson_and_alfredsson} focused specifically on the case $\alpha = 1$, which is unstable for the Poiseuille flow ($R_v = 0$), and showed that it experienced stabilization followed by destabilization for small to intermediate $R_v$, see Figure \ref{fig:modal_growth_rates}. Subsequently, at the so-called ``Branch-I" $R_v$ ($\approx 42.91$), exponential growth could be re-established (cf. \cite{fransson_and_alfredsson}, the mode became unstable ``again") and sustained until the ``Branch-II" $R_v$ ($\approx 636.16$), which initiated yet another region of stability.

Evidently, as verified in Figure \ref{fig:modal_growth_rates}, this classification is not appropriate for all combinations of $\left(\alpha, Re\right)$. As a simple example, the wavenumber $\alpha=0.5$ does not exhibit instability at $Re = 6000$ for the Poiseuille flow, rendering the notion of a Branch-I $R_v$ inherently ill-defined. Indeed, since the upper and lower branches of the Poiseuille neutral curve decay as $Re^{-1/11}$ and $Re^{-1/7}$ at large $Re$ (see, for example, \cite{lin1946stability, Cowley_Smith_1985}), such a $R_v$ cannot be demarcated for \textit{any} $Re$ at even smaller wavelengths, say $\alpha \geq 1.5$. \cite{GUHA_2010} recognized this and circumvented the problem by introducing a third branch (which marked the transition from unstable to stable when $R_v$ initially increases from zero) whose existence was predicated on the stability of $\left(\alpha, Re, R_v=0\right)$. In summary, a more refined characterization was needed, and here we attempt to add to the discussion.

\begin{figure}
\centering
\includegraphics[width=0.95\textwidth]{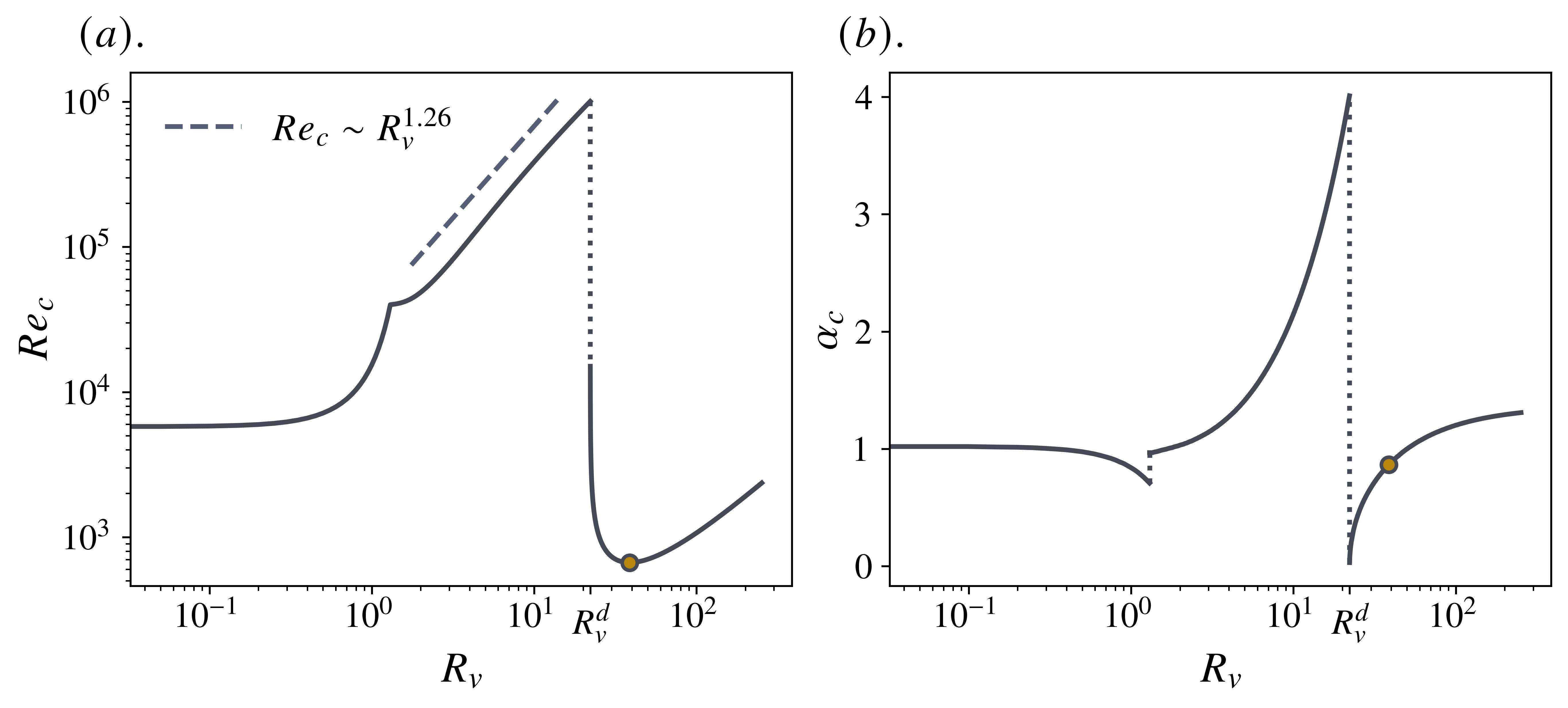}
\caption{The critical flow parameters versus the crossflow Reynolds number $R_v$; $(a)$, the critical Reynolds number $Re_c$, and $(b)$, the critical streamwise wavenumber, $\alpha_c$. In each panel, the dotted lines elucidate points of discontinuity, with $R_v = R_v^d$ being common to both $Re_c$ and $\alpha_c$. Furthermore, a circle indicates the flow parameters minimizing $Re_c$. Finally, in $(a)$, the dashed line represents the scaling $Re_c\sim R_v^{1.26}$.}
\label{fig:critical_parameters}
\end{figure}

Figure \ref{fig:critical_parameters} summarizes a numerical search for the critical parameters $\left(\alpha_c, Re_c\right)$ for the range $0\leq R_v\leq 250$. Similarly to previous studies, we ignore $R_v < 0$ due to the invariance of the eigenproblem under the transformation $\left(y, R_v\right)\to \left(-y, -R_v\right)$. Consistent with the narrative suggested by Figure \ref{fig:modal_growth_rates}, we observe a rather sharp, monotonic increase in the critical Reynolds number $Re_c$ through $R_v\lessapprox 1$. This trend persists beyond this interval, adjusting, however, to an approximate power-law behavior, $Re_c\sim R_v^{1.26}$. Throughout the latter range, the value of $Re_c$ increases rapidly to $Re_c\approx \mathcal{O}\left(10^6\right)$, before abruptly descending to $Re_c\approx \mathcal{O}\left(10^3\right)$ at $R_v=R_v^d \approx 22.175$. Shortly thereafter, the minimum of $Re_c\approx 667.48730$ is achieved at the turning point $R_v\approx 38.75$, which is in good agreement with the findings of \cite{fransson_and_alfredsson}. Meanwhile, the critical streamwise wavenumber $\alpha_c$ initially experiences a brief drop from its value at $R_v = 0$ ($\alpha_c\approx 1.02$), before recovering at $R_v\approx 1$ -- note that such a discontinuity does not exist for $Re_c$. Following this, with stronger crossflows, we observe an increasing preference for short-wavelength instabilities up to the critical value of $R_v = R_v^d$. Here, in conjunction with $Re_c$, another discontinuity is encountered and $\alpha_c$ rapidly descends to near-zero ($\approx \mathcal{O}\left(10^{-2}\right)$ with the extent of our computation) before rising to what appears to be an asymptote. We remark here that for $20.8\lessapprox R_v\lessapprox 22$, \cite{GUHA_2010} report an unconditional linear \textit{stability} for the crossflow-laden Poiseuille flow (in fact, even for the Couette-Poiseuille flow $\xi \neq 0$ -- see Section \ref{sec:wall_velocity}), which is in contrast to Figure \ref{fig:critical_parameters}. According to our understanding, this discrepancy appears to be due to a combination of insufficient numerical resolution (they used $N = 120$ collocation points) and a restriction of their search space to $Re \leq 10^6$. We verified that our calculations remained robust even when the resolution was doubled, indicating genuine instability. 

\begin{figure}
\centering
\includegraphics[width=\textwidth]{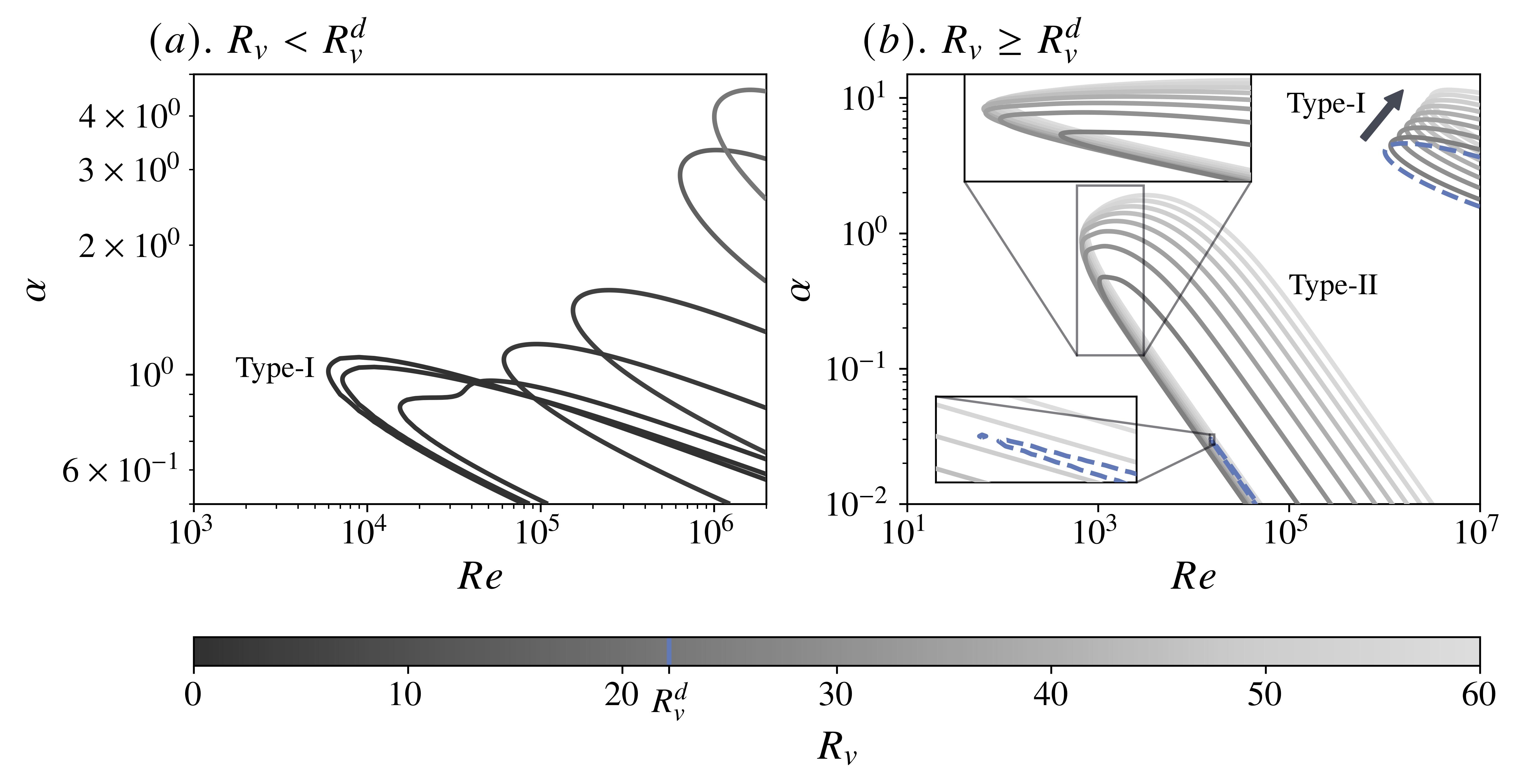}
\caption{Movement of the neutral curves in the $\left(\alpha, Re\right)$-plane for $(a)$, $R_v < R_v^d$ and $(b)$, $R_v \geq R_v^d$. Where appropriate, Type-I and Type-II instabilities have been labeled. For $(b)$, specifically, an inset zooms in on the development of the Type-II NSCs; in the same panel, the dashed contours correspond to $R_v = R_v^d$.}
\label{fig:neutral_curves_movement}
\end{figure}

To unravel the discontinuous nature of the critical parameters, Figure \ref{fig:neutral_curves_movement} details the movement of the neutral stability curves (NSCs) in the $\left(\alpha, Re\right)$-plane, particularly before and after $R_v = R_v^d$. Focusing first on $R_v<R_v^d$, Figure \ref{fig:critical_parameters}$(a)$ indicates that the presence of crossflow is exclusively stabilizing, and this is verified in Figure \ref{fig:neutral_curves_movement}$(a)$ by a net displacement of the NSCs in the direction of increasing $Re$. A secondary minimum develops at intermediate $R_v$ in this range, which quickly coalesces with the primary one and appears to signal a small jump in the NSC towards larger wavenumbers, consistent with the first discontinuity for $\alpha_c$ observed in Figure \ref{fig:critical_parameters}$(b)$. Henceforth, the NSCs continue to shift deeper into the upper-right corner of the $\left(\alpha, Re\right)$-plane.

Beyond $R_v = R_v^d$, more dynamic behavior can be resolved. Specifically, two \textit{different} NSCs begin to co-exist. The first set of unstable modes, relegated to $Re \geq O\left(10^6\right)$, can effectively be traced back to the (sole) NSC that occurs for $R_v <  R_v^d$ and, as such, can be interpreted as its continuation into this $R_v$ regime. Neither \cite{fransson_and_alfredsson} nor \cite{GUHA_2010} reported this, instead implying the presence of a single NSC for all unstable $R_v$; we label these (and their counterparts in  $R_v<R_v^d$) as ``Type-I" instabilities and note that in both $R_v$ regimes, they generally exhibit a monotonic displacement toward increasing $Re$ as the crossflow becomes stronger. The second group of instabilities, deemed ``Type-II" and active only in $R_v\geq R_v^d$, manifest in the form of an NSC emerging from $\alpha\approx 0$ and are, of course, precisely those reported in previous work since they formally define criticality for this $R_v$ regime. As highlighted in the inset of Figure \ref{fig:neutral_curves_movement}$(b)$ these modes temporarily destabilize before stabilizing, defining in the process the turning point observed for $Re_c$ in Figure \ref{fig:critical_parameters}$(a)$.

To dissect the physical mechanism driving the instability, one can turn to the perturbation energy budget. By writing the energy density as $u_ku_k/2$, where a repeated index implies Einstein summation, the total disturbance energy can be written as
\begin{equation}
    \mathsf{E} = \dfrac{1}{2}\int_V u_ku_k\,\mathrm{d}V
\end{equation}
where $V \equiv \left[0, 2\pi/\alpha\right]\times\left[-1,1\right]\times\left[0,2\pi/\beta\right]$ is taken as one full disturbance wavelength. Using the normal mode ansatz, Equation (\ref{eqn:normal_modes}), evolution equations for $\mathsf{E}$ then take the form 
\begin{equation}
\label{eqn:energy_budget}
    \dfrac{\mathrm{d}\mathsf{E}}{\mathrm{d}t} = \left\langle \mathsf{PR}\right\rangle - \left\langle\mathsf{VD}\right\rangle
\end{equation}
where
\begin{equation}
     \left\langle f\left(y\right) \right\rangle = \int_{-1}^1f\left(y\right)\,\mathrm{d}y
\end{equation}
Here, $\mathsf{PR}$ represents production against the background shear (contributed to only by $\mathsf{D}U$) and is responsible for the transfer of energy from the base flow to the disturbance through the action of the Reynolds stress. The second term $\mathsf{VD} \geq 0$ instead denotes viscous dissipation. In general, a (positive) production destabilizes, whereas dissipation stabilizes the disturbance field, although for a marginally stable mode, as we will discuss here, these contributions must exactly cancel.

\begin{figure}[t!]
\centering
\includegraphics[width=\textwidth]{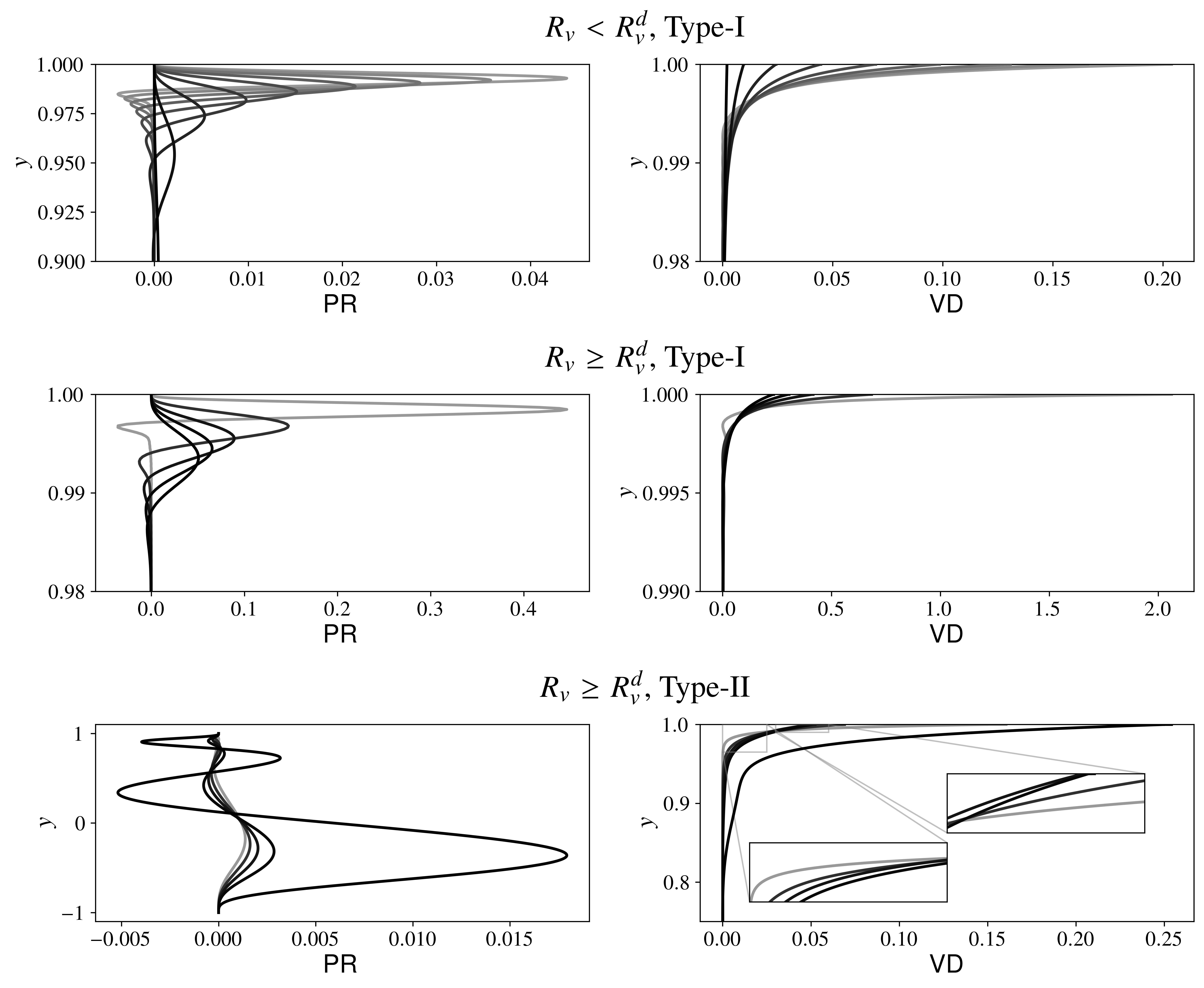}
\caption{Distributions of the energy production $\mathsf{PR}$ (left) and dissipation $\mathsf{VD}$ (right) for Type-I and Type-II modes at criticality. Here, for $R_v < R_v^d$, we consider $R_v\in\left\{0.5, 2.5, 5, 7.5, 10, 12.5, 15, 17.5, 20\right\}$, whereas for $R_v \geq R_v^d$, $R_v\in\left\{22.24, 26, 32, 45, 95\right\}$. Darker to lighter shades represent stronger crossflows in the labeled regimes.}
\label{fig:modal_analysis_energy_budget}
\end{figure}

Figure \ref{fig:modal_analysis_energy_budget} explores for Type-I and Type-II instabilities the spatial variation of terms that contribute to the energy budget at criticality. For Type-I modes below $R_v^d$, energy production $\mathsf{PR}$ operates primarily near the suction boundary, its peak becoming sharper in tandem with $R_v$. For the weakest crossflow strengths in this range, an additional small positive hump (not shown here) can also be resolved near the lower wall, although it decays very rapidly as $R_v$ increases. Similarly, viscous dissipation $\mathsf{VD}$ remains confined to the upper wall and also increases in conjunction with $\mathsf{PR}$, as required to ensure neutral growth. As highlighted in the second row of Figure \ref{fig:modal_analysis_energy_budget}, these trends appear to translate well to a Type-I criticality in $R_v\geq R_v^d$, which provides further evidence that they are, in fact, a continuation of Type-I modes from $R_v< R_v^d$. However, for Type-II instabilities, the picture changes dramatically. Here, $\mathsf{PR}$ develops noticeable regions of energy negation and, more importantly, a wide positive peak in the lower (injection) half of the channel. We find the latter behavior quite intriguing, particularly in light of recent work showing that it is, in fact, the boundary inflow (outflow) that supports destabilization (stabilization); see \cite{Kerswell_2015}. Equally important to mention here is that for moderate to large crossflows (such as those found in $R_v\geq R_v^d$), variations in the streamwise shear $\mathsf{D}U$ are primarily located near the upper wall (with $\mathsf{D}U\to 1/2$, the Couette value, elsewhere). However, since the amplitude of positive production instead peaks specifically near the lower wall, we conclude that the crossflow influences Type-II modes predominantly through modification of the Reynolds stress distribution.

Formally, the last row in Figure \ref{fig:modal_analysis_energy_budget} corresponds to the regime explored in Section 5 of \cite{GUHA_2010}. In their Figure 17, for example, they reported that the dissipation at criticality for $\left(\xi, R_v\right) = \left(0.5, 25\right)$ was completely overshadowed by the energy production throughout the channel width. Therefore, any removal of energy from the perturbation field arose directly from energy negation, rendering the movement of the critical layers, as they say, ``irrelevant''. Clearly, as highlighted in Figure \ref{fig:modal_analysis_energy_budget}, this is not an entirely robust mechanism, at least for $\xi = 0$, since, despite neutrality, $\mathsf{VD}$ for $R_v\geq R_v^d$ operates at least one order of magnitude higher than $\mathsf{PR}$ (though only very close to the wall) and generally intensifies with $R_v$. In other words, investigating the critical layers, the set of points where $U\left(y\right) = c_r$, could be instructive here. These points constitute regular singularities for the stability equations in the Rayleigh limit, but are smoothed over with the addition of viscosity in the Orr-Sommerfeld-Squire formulation. More importantly, it is well known that peaks in the amplitude of energy production are usually located within these layers \citep{mack_BL_stability, schmidstability, GUHA_2010, McKEON_SHARMA_2010}.
\begin{figure}
\includegraphics[width=\textwidth]{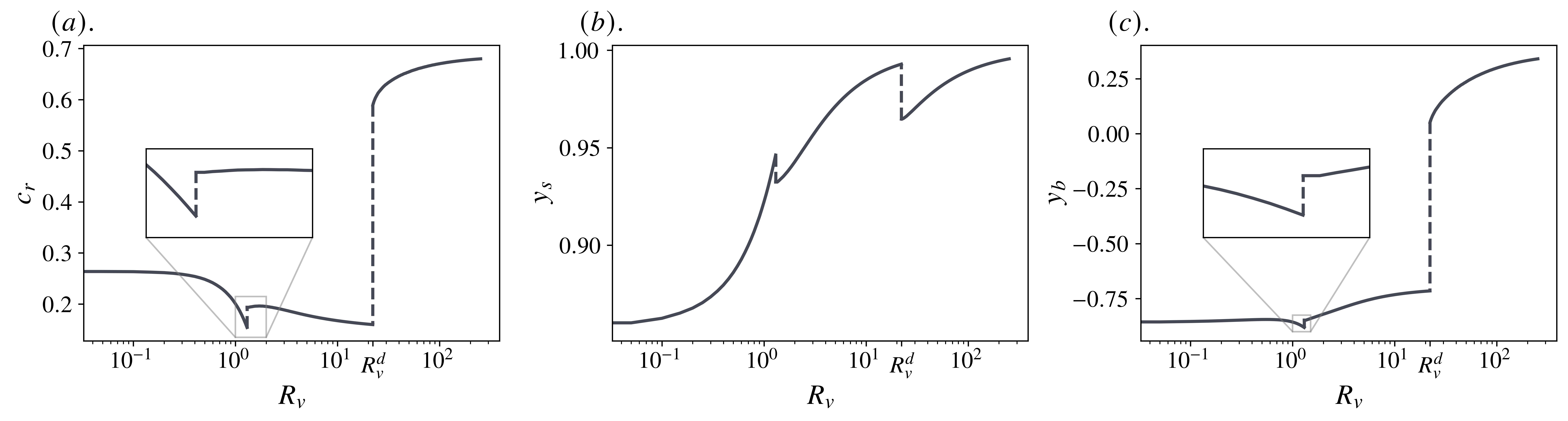}
\caption{Movement of the critical layers with $R_v$; $(a)$, the real part of the phase speed at criticality (note that $c_i = 0$ by definition of neutral stability), $(b)$, $y_s$, the critical layer near the suction (upper) wall, and $(c)$, $y_b$, the critical layer near the blowing (lower) wall.}
\label{fig:critical_layers}
\end{figure}

Figure \ref{fig:critical_layers}$(a)$ presents the variation of $c_r$ at criticality versus $R_v$; we immediately observe discontinuities synonymous with those of $\alpha_c$, Figure \ref{fig:critical_parameters}$(b)$. The exact location of the associated critical layers can be explicitly determined as the solution(s) to the following transcendental equation
\begin{equation}
\label{eqn:critical_layers_eq}
    U\left(y\right) = c_r \implies y - \csch R_v e^{R_v y} = \Pi_1,
\end{equation}
where we have elected to set
\begin{align}
    \Pi_1 & = \dfrac{\Pi_2c_r}{R_v} - \coth R_v ,\\
    \Pi_2 & = R_v\coth R_v - 1 - \log\left(R_v\csch R_v\right).
\end{align}
Equation (\ref{eqn:critical_layers_eq}) is satisfied by
\begin{equation}
    y = \Pi_1 - \dfrac{1}{R_v}W_n(\Pi_3).
\end{equation}
where $W_n$ is the $n^{\mathrm{th}}$-branch of Lambert's $W$-function and $\Pi_3 = -R_v\csch R_v e^{\Pi_2R_v}$. Here, since $-1/e\leq \Pi_3<0$, there exist two distinct critical layers (which can alternatively be concluded by noting that $c_r < 1$ and inspecting the velocity profiles in Figure \ref{fig:geometry}$(b)$), and we only need to consider $n\in\left\{-1, 0\right\}$. We observe that, as $R_v\to 0$, the symmetry of the resulting profile requires the absolute values of these roots to coalesce, despite being in opposite halves of the channel. Thus, the solutions, which we label $y_s$ and $y_b$, respectively, can be naturally identified with the suction (top) and blowing (lower) boundaries. On the other hand, due to the asymmetry inflicted upon $U$ by the crossflow, one can expect both critical layers to eventually shift toward the upper half of the channel, an intuition that is verified in Figures \ref{fig:critical_layers}$(b$--$c)$. Specifically, commensurate with the movement of the peak in energy production for Type-I modes, $y_s$ generally increases, effectively approaching $y_s = 1$ for sufficiently large $R_v$. Separately, beyond $R_v=R_v^d$, $y_b$ undergoes a sudden jump into the suction half of the channel, which is consistent with the wide production peaks (spanning almost the entire lower half-width) observed for Type-II modes in the last row of Figure \ref{fig:modal_analysis_energy_budget}.

\section{\label{sec:nonmodal_analysis} Non-Modal Analysis}

We now direct our focus toward non-modal energy amplification. As shown in Section \ref{sec:modal_analysis}, an increase in the crossflow component starting from $R_v=0$ generally has a stabilizing effect, at least in the sense of the critical Reynolds number. Following a paradigm shift
at $R_v=R_v^d$, however, positive growth rates can be achieved for Reynolds numbers as low as $Re\approx700$. When operating solely on this information, one would conclude that the most optimal (and practical) stabilization is granted by weak to, at best, modest $R_v$. In this section, we show that non-normality indicates an entirely different story.

\begin{figure}[t!]
\centering
\includegraphics[width=\textwidth]{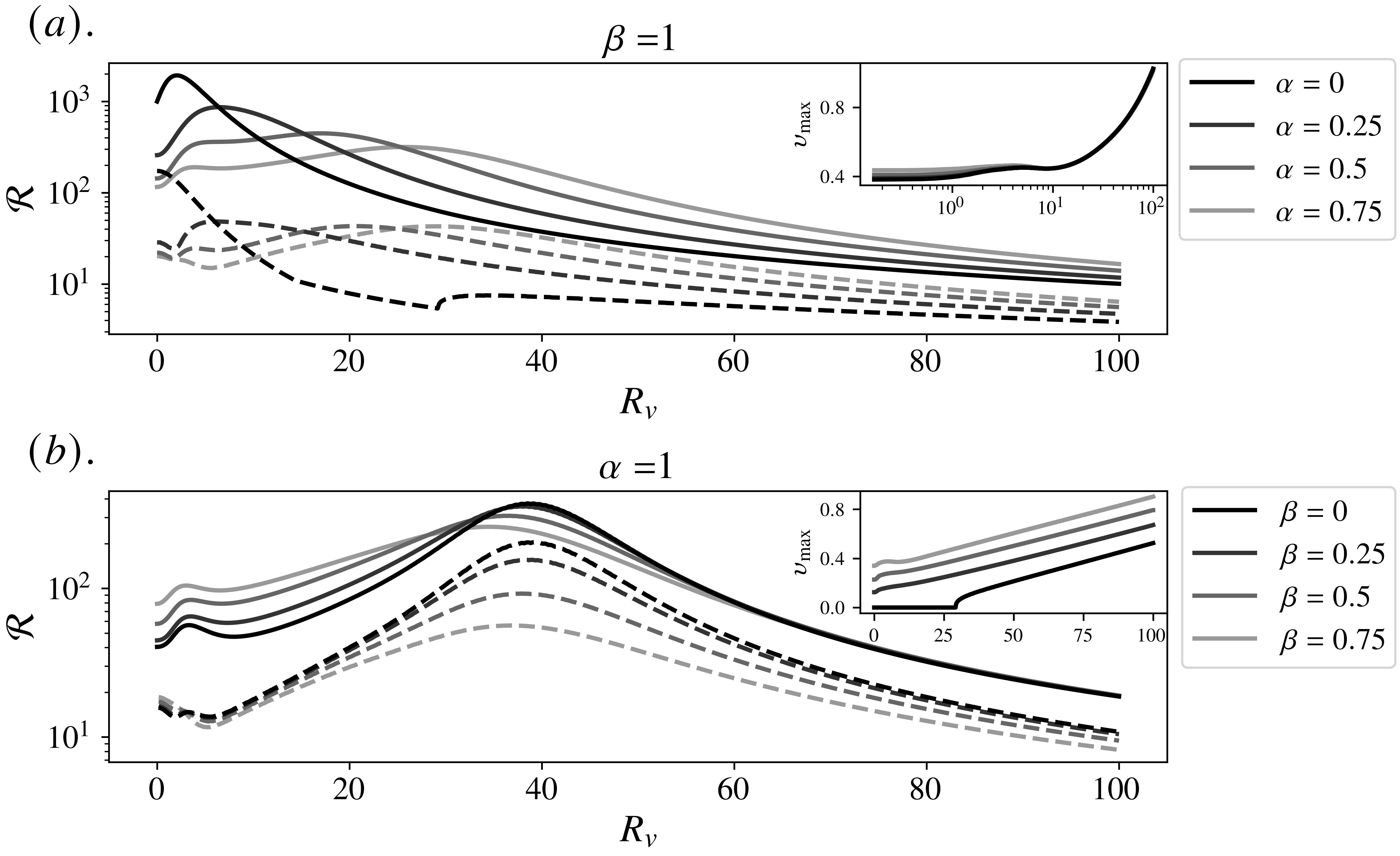}
\caption{For various wavenumber pairs at $Re = 500$, plots of $\mathcal{R}$, the maximum resolvent norm over all real forcing frequencies; $(a)$, $\beta = 1$ and $\alpha$ increased from $\alpha = 0$ to $\alpha=0.75$ in increments of 0.25 and $(b)$, $\alpha = 1$, with $\beta$ varied in a similar manner. The insets show the maximizing frequency $\upsilon_{\max}$ and, for each wavenumber combination, an associated dashed line represents the lower bound of Equation (\ref{eqn:resolvent_bounds}) evaluated at $\upsilon = \upsilon_{\max}$.}
\label{fig:resolvent_norms}
\end{figure}

We begin with the resolvent $\mathsf{R}$. In particular, the following bounds on the norm $\left\lVert \mathsf{R}\right\rVert_E$ are standard
\begin{equation}
\label{eqn:resolvent_bounds}
    \dfrac{1}{\mathrm{dist}\left(\upsilon, \Lambda\left(\mathsf{S}\right)\right)} \leq \left\lVert \mathsf{R}\right\rVert_E \leq \dfrac{\kappa\left(\mathsf{W}\right)}{\mathrm{dist}\left(\upsilon, \Lambda\left(\mathsf{S}\right)\right)}
\end{equation}
where $\mathrm{dist}\left(\upsilon, \Lambda\left(\mathsf{S}\right)\right)$ is the shortest distance between $\upsilon$ and the spectrum $\Lambda\left(\mathsf{S}\right)$ (equivalent to the spectral radius of $\mathsf{R}$) and  $\kappa\left(\mathsf{W}\right)$ is the 2-norm condition number of $\mathsf{W}$ \citep{tref_pseudospec, reddyschmidhenn, schmidstability}. For systems governed by normal operators, $\kappa = 1$, so that the bounds in Equation (\ref{eqn:resolvent_bounds}) coalesce into equality. On the other hand, for non-normal linear dynamics, $\kappa\gg 1$ and the eigenfunctions of the underlying operator can be highly oblique, so that even pseudoresonant $\upsilon$ far from an eigen-frequency are capable of generating a substantial response. Thus, to probe the potential for this energy growth, we define the following quantity
\begin{equation}
\label{eqn:hinf_norm}
    \mathcal{R}\left(\alpha, \beta, Re, R_v\right) = \max_{\upsilon\in\mathbb{R}}\left\lVert \mathsf{R}\right\rVert_E\left(\alpha,\beta, Re, R_v, \upsilon\right)
\end{equation}
which, if the resolvent is interpreted as a transfer function between the excitation and its
response, represents an $H_\infty$-norm. In Figure \ref{fig:resolvent_norms}, we visualize $\mathcal{R}$, along with the maximizing frequency $\upsilon = \upsilon_{\max}$, for some sample wavenumbers at $Re = 500$, which is sub-critical and, therefore, admits no modal instability for all $R_v$ treated here. The dashed lines denote the lower bound in Equation (\ref{eqn:resolvent_bounds}), plotted for $\upsilon_{\max}$; the significant discrepancy observed with $\mathcal{R}$ is entirely a consequence of non-normality.

We first comment on longitudinal ($\alpha = 0$) and transverse ($\beta = 0$) perturbations. The former class typically elicits the strongest resolvent response for purely streamwise base flows. Therefore, we expect this trend to persist at least for weak $R_v$, and this is verified in Figure \ref{fig:resolvent_norms}$(a)$. In fact, we see that the amplification of streamwise-independent modes can even be intensified in the presence of small amounts of crossflow. As an example, for $R_v\approx 2$, for which Figure \ref{fig:critical_parameters}$(a)$ predicts $Re_c\approx 48500$, $\mathcal{R}\approx 2000$, so that a unit norm forcing can produce $\mathcal{O}(10^3)$ energy growth (compare this with $\mathcal{R}\approx 1000$ at $R_v = 0$). This is the first indication of the unreliability of eigenvalues; when the effects of non-normality are taken into account, weak crossflows are, in fact, the most prone to excitation.

Interestingly, for relatively large $R_v$, Figure \ref{fig:resolvent_norms}$(a)$ indicates that $\mathcal{R}$ tends to decay for longitudinal perturbations; in contrast; an increasingly stronger response from spanwise-independent modes is seen in Figure \ref{fig:resolvent_norms}$(b)$, a gap that appears to peak at $R_v\approx 40$. More generally, in intermediate $R_v$ regimes, oblique disturbances constitute the main preference. Compared to the crossflow-independent case, these trends are inherently distinct but not unexpected, since the addition of a third-order inertial term in the Orr-Sommerfeld-Squire system fundamentally alters its anatomy and, therefore, that of the underlying non-normality (see Appendix \ref{sec:appendix}). However, what is crucial to note here is that $\mathcal{R}$ is only truly attenuated when $R_v$ becomes very large, say $R_v\geq 70$. In other words, it is, in fact, the weakest crossflows that instigate the strongest linear growth, possibly transition arising from subsequent non-linear processes. Given that most practical applications are typically noise-heavy and that such (relatively) strong crossflows are often not even feasible, we begin to see that crossflow-based modulation in channels might not be as appropriate a choice as suggested in the previous literature.

\begin{figure}[t!]
\includegraphics[width=\textwidth]{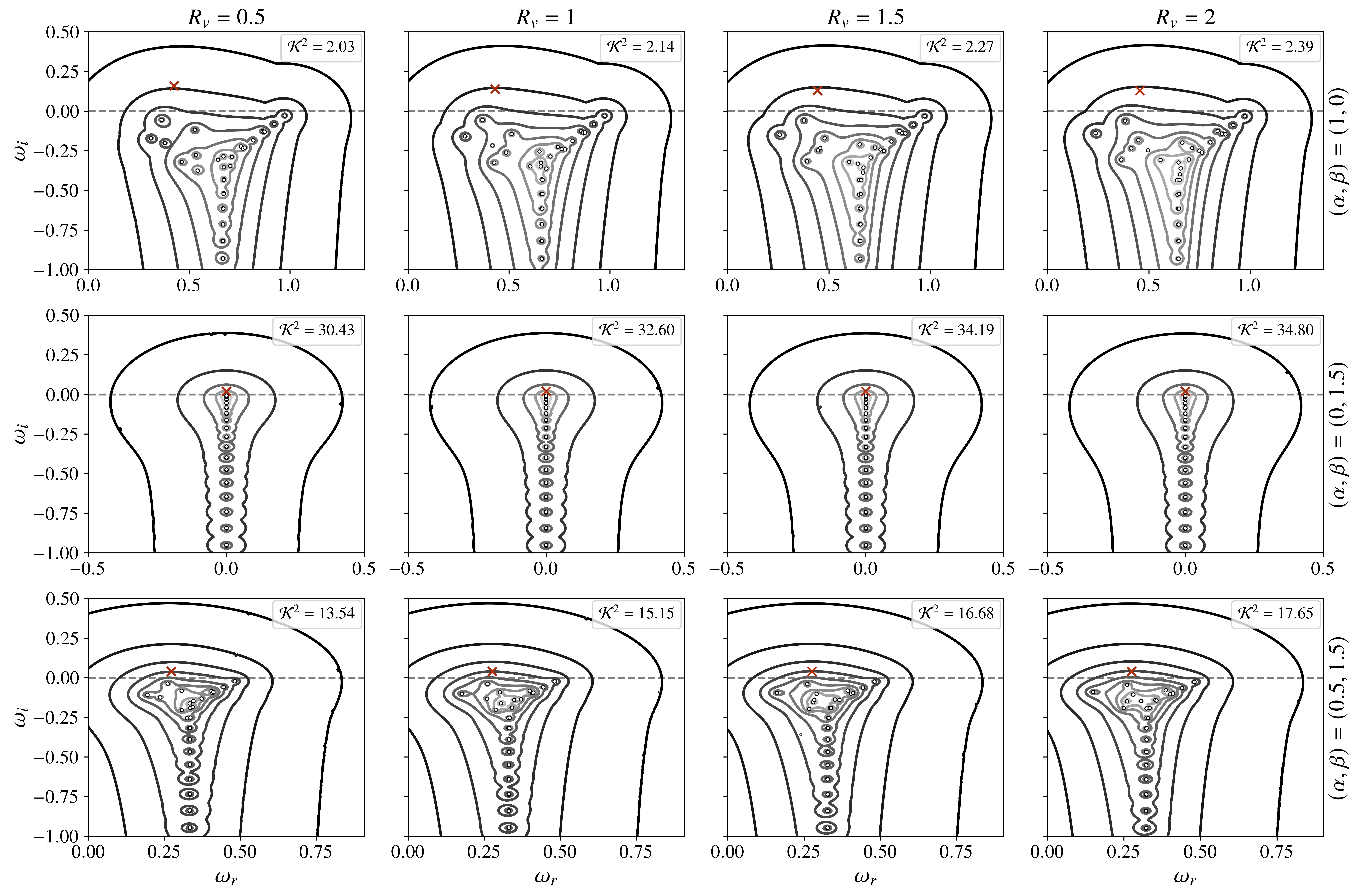}
\caption{Pseudospectral contours, $\log\,\,\lVert \mathsf{R}^{-1}\rVert_{-E}$, calculated at $R_v = 0.5, 1, 1.5, 2$ (left-to-right) for some representative wavenumber pairs. Darker to lighter (or, alternatively, outer to inner) shades correspond to decreasing contour values, i.e., increasing amplification. Top row, $(\alpha, \beta) = (1, 0)$, with contours in each panel ranging from $-0.5$ to $-4.5$ in decrements of $-0.5$; middle row, $(\alpha, \beta) = (0, 1.5)$, with contours ranging from $-0.6$ to $-3$ in decrements of $-0.6$; and bottom row, $(\alpha, \beta) = (0.5, 1.5)$, with contours ranging from $-0.5$ to $-4.5$ in decrements of $-0.5$. The associated spectra (in terms of $\omega$) have been denoted by circles. A red cross marks the $\upsilon\in\mathbb{C}$ maximizing $\mathcal{K}$, the latter quantity displayed in the upper right corner of each panel.}
\label{fig:pseudospectra}
\end{figure}

\begin{figure}[t!]
\centering
\includegraphics[width=\textwidth]{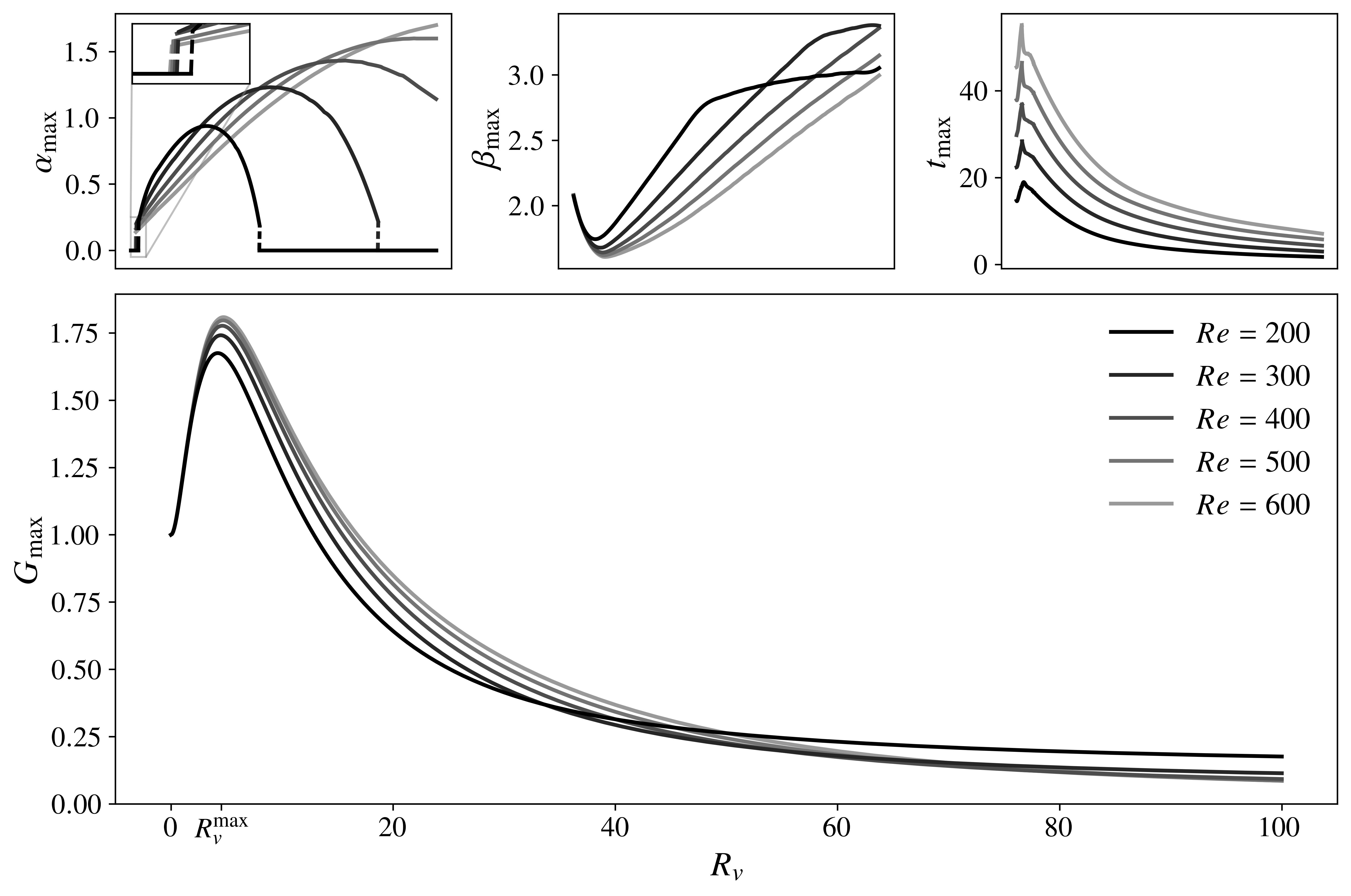}
\caption{Plots of $G_{\max}$ versus $R_v$, normalized against $G_{\max}\left(Re,R_v = 0\right)$, for $Re = 200$ to $Re = 600$ in increments of 100. $R_v = R_v^{\max}$, marked appropriately, denotes the crossflow strength that suffers the largest algebraic growth. Top: the wavenumbers $\left(\alpha_{\max}, \beta_{\max}\right)$ and the time $t = t_{\max}$ corresponding to $G_{\max}$.}
\label{fig:gmax}
\end{figure}

Moving to the complex plane, $\upsilon\in\mathbb{C}$, in Figure \ref{fig:pseudospectra}, we illustrate, as is customary, the contours of $\lVert \mathsf{R}^{-1}\rVert_{-E}$, where
\begin{equation}
    \lVert \mathsf{R}^{-1}\rVert_{-E} = \sigma_{\min}(\mathsf{W}\mathsf{R}^{-1}\mathsf{W}^{-1})
\end{equation}
and $\sigma_{\min}$ represents the minimum singular value of the operator $\mathsf{W}\mathsf{R}^{-1}\mathsf{W}^{-1}$. To interpret this in the context of the $\epsilon$-pseudospectra, we note that $\left\lVert \mathsf{R}\right\rVert_{E} = \lVert \mathsf{R}^{-1}\rVert_{-E}^{-1}$, so that
\begin{equation}
    \Lambda_\epsilon = \left\{\upsilon\in\mathbb{C} \colon \lVert \mathsf{R}^{-1}\rVert_{-E}\leq \epsilon\right\}
\end{equation}
Thus, within the level set $\lVert \mathsf{R}^{-1}\rVert_{-E} = \epsilon$, amplification of the order of $1/\epsilon$ can be induced as a consequence of harmonic forcing. The pseudospectra are extremely informative, revealing crucial insight on the sensitivity of the spectrum to operator-level perturbations (see \cite{trefethen1997bau} and \cite{symon}) or, as is more relevant here, energy growth in the unforced initial-value problem. More specifically, the Hille-Yosida Theorem states that $G\leq 1$ if and only if the pseudospectra lie sufficiently close to the lower (stable) half-plane \citep{reddyschmidhenn}. This condition can be made explicit through either the pseudospectral abscissa or, alternatively, the Kreiss constant $\mathcal{K}$, defined as follows
\begin{equation}
    \mathcal{K} = \sup_{\mathfrak{I}\left(\upsilon\right) > 0} \mathfrak{I}\left(\upsilon\right)\left\lVert \mathsf{R}\right\rVert_E
\end{equation}
which in turn provides the following lower bound
\begin{equation}
    \max_{t>0}G\left(\alpha,\beta, Re, R_v,t\right) \geq \mathcal{K}^2
\end{equation}
Hence, substantial transient growth can be achieved if the pseudospectra protrude significantly into the unstable half-plane. Returning to Figure \ref{fig:pseudospectra}, we immediately observe strong pseudo-resonance down to $\epsilon\approx 10^{-5}$, indicative yet again of the strong non-normality pervading the linear dynamics of this system. Reminiscent of the trends shown in Figure \ref{fig:resolvent_norms}, this amplification evidently worsens in tandem with the crossflow strength, eventually decaying only in the intermediate to large $R_v$ range (which is not shown here). Variations in the Kreiss constant suggest that, at least for weak crossflows, streamwise-independent disturbances remain the most relevant for the unforced problem, although oblique modes appear to be not too far behind. This is indeed representative of the ground truth, as we now proceed to discuss.

In particular, for various $Re$, Figure \ref{fig:gmax} summarizes a numerical sweep for $G_{\max}$, defined as the maximum admissible transient gain in energy over time and across wavenumber space, i.e.
\begin{equation}
    G_{\max}\left(Re, R_v\right) = \max_{\alpha,\beta, t} G\left(\alpha, \beta, Re, R_v, t\right)
\end{equation}
Here, to concentrate specifically on amplification that occurs in the absence of exponential instability, we have limited our analysis to Reynolds numbers that are globally sub-critical, $Re\lessapprox 650$. In doing so, we readily observe a familiar pattern: for all $Re$ considered here, peak energy growth is only enhanced by small levels of crossflow, with $R_v = R_v^{\max}\approx 4.5$ consistently characterizing the worst-case scenario. Furthermore, it is precisely at very large $R_v$ that $G_{\max}$ begins to appreciably weaken relative to the Poiseuille flow. Note that for the asymptotic suction boundary layer, \cite{FRANSSON2003259} reported a similarly declining -- but still significant -- amplification relative to the Blasius (no suction) case, although this conclusion could very well be specific to the Reynolds number they opted to focus on. 

Regardless, as predicted by Figure \ref{fig:pseudospectra}, the optimal gain is indeed achieved for streamwise-elongated perturbations for small $R_v$ (say, $R_v \leq 2$), after which oblique modes become the norm (note that \cite{Chen_2021} apparently did not see the former for $Re = 1000$). Interestingly, however, at the lower end of the Reynolds numbers treated here, sufficiently strong crossflows can once again establish a preference for streamwise-independent disturbances; we conjecture that expanding the range of $R_v$ investigated might lead to similar behavior at larger $Re$, although we did not attempt to verify this. Finally, the optimal time $t_{\max}$ also temporarily increases and then decreases monotonically in conjunction with $R_v$ (equivalently $G_{\max}$), although its peak was found instead at $R_v \approx 2 \neq R_v^{\max}$. The immediate implication is that algebraic growth operates on much longer timescales in the presence of crossflow. \cite{FRANSSON2003259} concluded similarly for ASBL but the effect is, in a sense, more pronounced here, since $G_{\max}$ also increases.

\begin{figure}[t!]
\centering
\includegraphics[width=\textwidth]{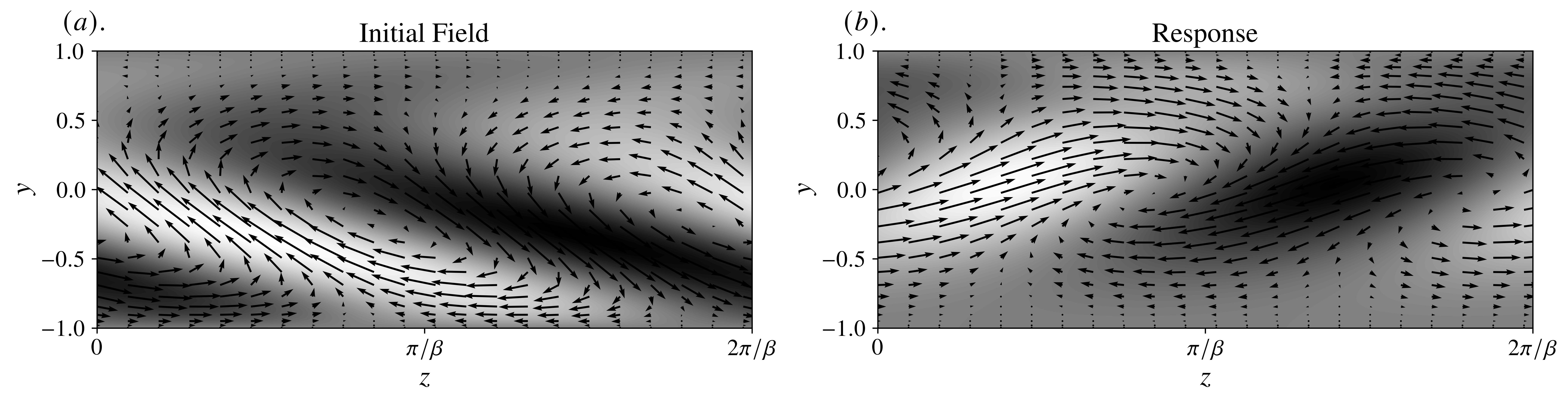}
\caption{Cross-stream ($y$--$z$) view of $(a)$, the initial perturbation and $(b)$, the response at optimal time associated with $G_{\max}$ for $R_v = R_v^{\max}$. Color and arrows represent, respectively, the streamwise and cross-stream components.}
\label{fig:optimal_perturbation}
\end{figure}

\begin{figure}
\centering
\includegraphics[width=\textwidth]{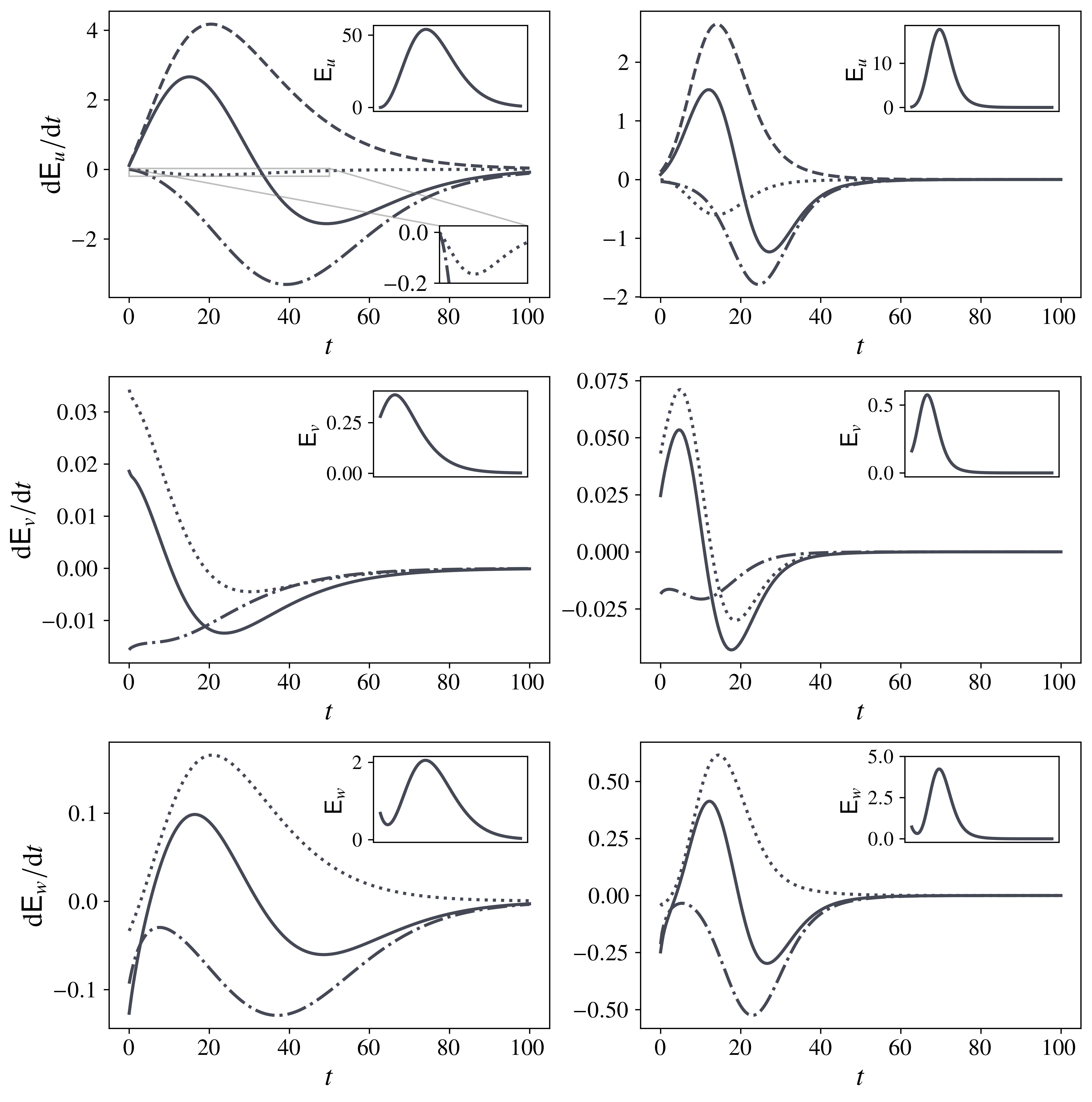}
\caption{Budgets for the perturbation energy contained in each velocity component; left column, $R_v = R_v^{\max}$ and right column, $R_v = 20$. In each panel, a dashed line indicates energy production, dotted, pressure redistribution, dash-dot, viscous dissipation, and solid, total. The insets show the variation of the perturbation energies $\mathsf{E}_u, \mathsf{E}_v$, and  $\mathsf{E}_w$ in time.}
\label{fig:stresses_energy_budget}
\end{figure}

Figure \ref{fig:optimal_perturbation} shows the initial condition and response pair attached to the optimal gain calculated for $R_v = R_v^{\max}$. Streaky structures, supported in both $x$ and $z$, develop at initial time and are then subsequently tilted and advected toward the suction boundary. A crucial question here is the precise mechanism driving the variation in the disturbance energy. It is well-known that longitudinal perturbations develop primarily via the lift-up effect, whereby strong streamwise streaks develop as a consequence of mean momentum transport in the wall-normal direction \citep{ellingsenpalm, landahl1, landahl2}. On the other hand, for perturbations with finite streamwise wavenumbers, the tilting mechanism of Orr dominates instead \citep{orr1907, orr_jiao}. In oblique disturbances, these processes operate simultaneously and \cite{Chen_2021} argued using a decomposition of the energy production proposed by \cite{production_decomposition} that at larger $R_v$, the Orr mechanism became increasingly relevant. When juxtaposed with Figure \ref{fig:gmax}, this result should not be too surprising, since $\alpha_{\max}$ generally also increases with $R_v$. To refine this approach, however, we consider the time variation of the perturbation energy contained in each velocity component
\begin{align}
\label{eqn:uu}
    \dfrac{\mathrm{d}\mathsf{E}_u}{\mathrm{d}t} & = \left\langle \mathsf{PR}_u\right\rangle + \left\langle \mathsf{PVC}_u\right\rangle + \left\langle \mathsf{VD}_u\right\rangle \\
    \label{eqn:vv}
    \dfrac{\mathrm{d}\mathsf{E}_v}{\mathrm{d}t} & = \left\langle \mathsf{PVC}_v\right\rangle + \left\langle \mathsf{VD}_v\right\rangle \\ 
    \label{eqn:ww}
    \dfrac{\mathrm{d}\mathsf{E}_w}{\mathrm{d}t} & = \left\langle \mathsf{PVC}_w\right\rangle + \left\langle \mathsf{VD}_w\right\rangle
\end{align}
where $\mathsf{PR}_i$ represents energy production, $\mathsf{PVC}_i = \mathsf{PS}_i + \mathsf{PD}_i$ the pressure-velocity correlation, $\mathsf{PS}_i$ the pressure-rate-of-strain, $\mathsf{PD}_i$ the pressure diffusion, and $\mathsf{VD}_i$ the viscous dissipation; see, for example, \cite{Hack_Zaki_2015}. One can further relate the terms in Equations (\ref{eqn:uu})--(\ref{eqn:ww}) to those in Equation (\ref{eqn:energy_budget}) as follows
\begin{align}
\mathsf{E}_u + \mathsf{E}_v + \mathsf{E}_w & = \mathsf{E} \\
    \left\langle \mathsf{PR}_u\right\rangle & = \left\langle \mathsf{PR}\right\rangle \\
    \left\langle \mathsf{PVC}_u\right\rangle + \left\langle \mathsf{PVC}_v\right\rangle + \left\langle \mathsf{PVC}_w\right\rangle & = 0 \\
    \left\langle \mathsf{VD}_u\right\rangle  + \left\langle \mathsf{VD}_v\right\rangle  + \left\langle \mathsf{VD}_w\right\rangle & = -\left\langle \mathsf{VD}\right\rangle 
\end{align}
From here, we note that when $\alpha = 0$, $\mathsf{PVC}_u$ vanishes, so that $\mathsf{E}_v$ and $\mathsf{E}_w$ decay simply due to a combination of the inter-variance pressure redistribution and viscous dissipation. As a result, we have $\mathsf{E}\approx \mathsf{E}_u$, whose growth in turn depends on the injection of energy from the mean shear through $\mathsf{PR}_u$. In other words, any changes in the amplification of streamwise-independent modes can be primarily attributed to modifications in the base shear and, by extension, to the Reynolds stresses. On the other hand, when $\beta=0$, the off-diagonal term in the Orr-Sommerfeld-Squire operator vanishes, so that the impact of non-normality is anyway diminished. Thus, we are left to deal primarily with oblique disturbances. In what follows, the pressure diffusion term $\mathsf{PD}_i$ was found to contribute negligibly to $\mathsf{PVC}_i$ and has therefore been omitted for the sake of clarity.

Using the energy-optimal initial conditions associated with $R_v = R_v^{\max}$ and $R_v = 20$ (note that these disturbances are both oblique), Figure \ref{fig:stresses_energy_budget} compares the development of the energy budget over time for each component of the velocity. The primary motivator for the discrepancy in $G_{\max}$ can be immediately identified as the significantly attenuated production term, which evidently tapers around $t = 45$ for $R_v = 20$. In contrast, $\left\langle\mathsf{PR}_u\right\rangle$ for $R_v = R_v^{\max}$ operates longer and achieves much larger amplitudes. Furthermore, pressure redistribution remains active in both cases, although its influence is much more pronounced for the sub-optimal $R_v = 20$. On inspection, this energy is transferred to both the wall-normal and the spanwise perturbations, and it is interesting to note that $\mathsf{E}_v$ ultimately achieves a slightly larger amplitude for $R_v = 20$, so that a stronger amplification of the streaks observed in Figure \ref{fig:optimal_perturbation} might be expected due to the lift-up process. However, we see that this effect is negated not only by a moderate increase in dissipation, but also by the rapid extraction of energy from $\mathsf{E}_v$ to $\mathsf{E}_w$ through a strongly negative pressure redistribution term starting from, say, $t \approx 12.5$. Consequently, any increase in the wall-normal perturbations is quickly eroded and energy production suffers. The net influx of energy into $\mathsf{E}_w$ appears to primarily dissipate away, in part, due to the lack of a feedback mechanism (e.g., a mean spanwise shear) to amplify it.

\section{\label{sec:wall_velocity} The Presence of Wall Velocity}

\begin{table}
\centering
\begin{tabular}{||c||c||c||c||c||}
\hline
 $R_v$ & $\xi$ &$Re_c$ $(\times10^6)$ & $\alpha_c$&$N_\mathrm{conv}$ \\\hline\hline
21.9 & 0.015 & 1.083334 & 3.956539 & 512 \\
 20.9 & 0.025 & 1.080632 & 3.956539 & 512 \\
 21.75 & 0.03 & 1.175339 & 4.144285 & 512 \\
 21 & 0.05 & 1.272604 & 4.037017 & 512 \\
 21.25 & 0.075 & 1.569059 & 4.340941 & 512 \\
 21.5 & 0.1 & 2.043112 & 4.546928 & 512 \\
 21.55 & 0.125 & 2.859688 & 5.225415 & 1024 \\
 21.3 & 0.16 & 6.950775 & 8.307361 & 2560 \\
 \hline
\end{tabular}
\caption{Flow parameters at criticality for select pairs $\left(\xi, R_v\right)$ such that $20.8\leq R_v\leq 22$; $N_\mathrm{conv}$ represents the resolution (number of Chebyshev modes) that was required for resolving the instability (i.e., that the instability remained robust to further increases in numerical fidelity).}
\label{table:xi_crit_params}
\end{table}

If either wall is translated in the streamwise direction with some finite velocity, say $U_c\neq 0
$, we recover the base flow of \cite{hains} and \cite{GUHA_2010}. More specifically, if this moving boundary is taken to be the upper (suction) one, the dimensionless streamwise velocity becomes
\begin{equation}
\label{eqn:pcp_base_flow}
    U\left(y, \xi, R_v\right) = \begin{cases}
        R_v\left(\xi R_v + 4y + \left(4-\xi R_v\right)\coth R_v-e^{R_v y}\left(4-\xi R_v\right)\csch R_v\right)/\left(\xi R_v^2 - 4 + \Xi_1\right), & \xi R_v \leq \Xi_2, \\[7.5pt]
        \left(\xi R_v + 4y + \left(4-\xi R_v\right)\coth R_v-e^{R_v y}\left(4-\xi R_v\right)\csch R_v\right)/2\xi R_v, & \xi R_v > \Xi_2. \\
    \end{cases}
\end{equation}
where $\xi = U_c/U_p$ and we have adopted a non-dimensionalization similar to that in Section \ref{ssec:base_velocity_profiles}. Furthermore, we have denoted
\begin{align}
    \Xi_1 & = R_v\left(4 - \xi R_v\right)\coth R_v - 4\log\left(R_v\left(4 - \xi R_v\right)\csch R_v/4\right),\\
    \Xi_2 & = 4\left(1 - \sinh R_v/R_v e^{R_v}\right)
\end{align}
We note two distinct scenarios in Equation (\ref{eqn:pcp_base_flow}), the need for which arises from the fact that both the crossflow and the Couette component skew the velocity profile towards the upper half of the channel \citep{GUHA_2010}. Thus, for strong enough crossflows (alternatively, large enough wall speeds), the maximum of the streamwise velocity will necessarily occur at the moving boundary, that is, $U_m = U_c$ and $y\left(U = U_m\right) = h$. If the pair $\left(\xi, R_v\right)$ is chosen such that $\xi R_v = 4 > \Xi_2$, one recovers $U\left(y, \xi, R_v\right) = y$, which constitutes the ``generalized'' Couette profile of \cite{nicoudandangilella}. Throughout this section, we limit our attention to wall speeds $\xi\in\left[0,1\right]$, as in the prior literature.

\subsection{Modal Instability Between $20.8\leq R_v\leq 22$ for $\xi > 0$}

Although not the primary focus in this work, we begin by reporting in Table \ref{table:xi_crit_params} samples from a set of novel instabilities that we were able to compute for crossflow-laden Couette-Poiseuille flow for $20.8\leq R_v\leq 22$. Previous work in this range by \cite{GUHA_2010} posited unconditional linear \textit{stability} for all $\xi \geq 0$. However, as demonstrated in Section \ref{sec:modal_analysis} and as is now apparent here, this is clearly an inaccurate conclusion arising again from a combination of a lack of polynomial resolution and a truncated search region. Despite this, we determined that the instability was short-wavelength, roughly consistent with the trends they calculated as $R_v \to 20.8$ (cf. Figure 20 in their paper). Although we did not conduct a detailed investigation to verify this, our numerical experiments suggest that the bounds offered in the stability phase diagram of \cite{GUHA_2010} might not be robust and could potentially be tightened if spatial discretization is refined.

\subsection{Transient Growth}

\begin{figure}[t!]
\centering
\includegraphics[width=\textwidth]{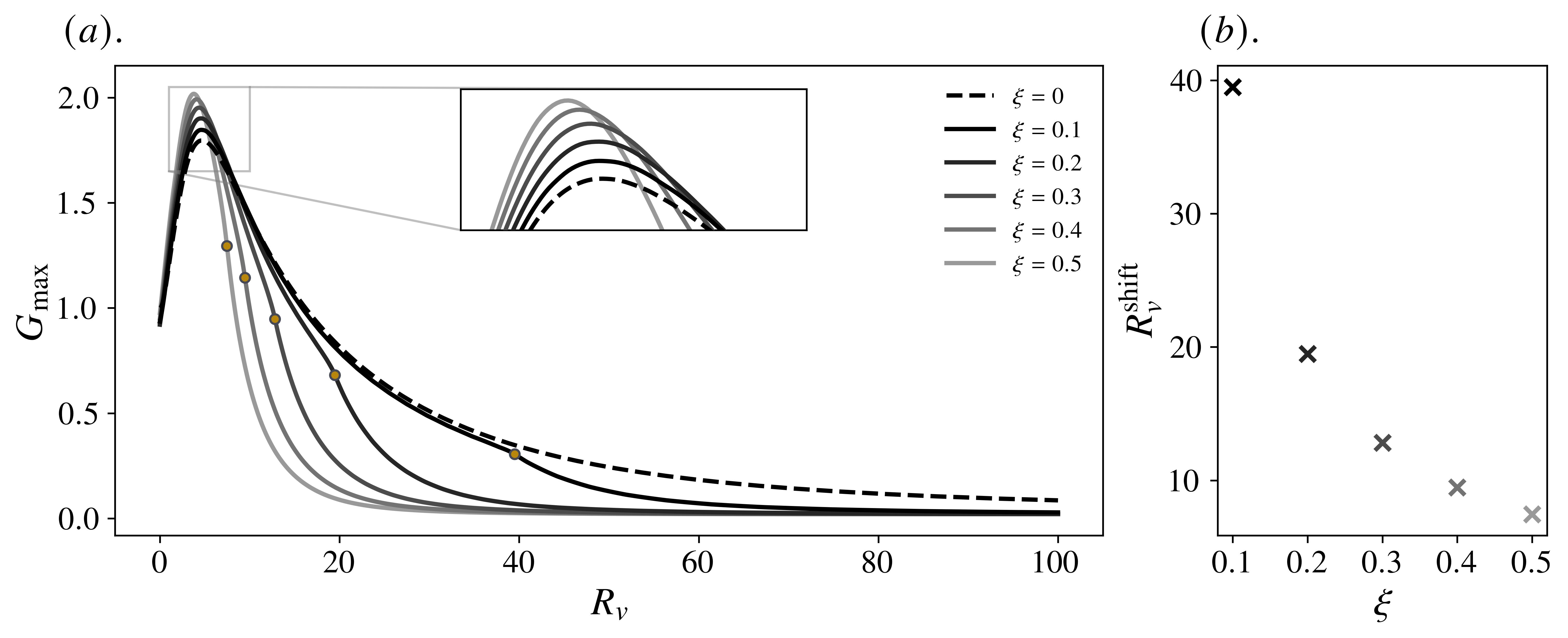}
\caption{$(a)$, plots of $G_{\max}$, the maximal energy gain, in the presence of wall motion versus $R_v$ for $\xi > 0$ and $Re = 500$; the reference case, $\xi = 0$, is depicted with a dashed line. All curves have been normalized by $G_{\max}$ at this $Re$ for Poiseuille flow, $\xi = 0 = R_v$. The yellow circles depict the minimal crossflow Reynolds number $R_v^\mathrm{shift}$ such that $\xi R_v > \Xi_2$, cf. Equation (\ref{eqn:pcp_base_flow}); these have been separately visualized in $(b)$.}
\label{fig:pcp_gmax}
\end{figure}

For various choices of the non-dimensional wall speed $\xi$, Figure \ref{fig:pcp_gmax}$(a)$ illustrates how the optimal gain $G_{\max}$ is modified in the presence of both crossflow and wall motion at $Re = 500$ (which remains globally sub-critical for $\xi > 0$). Two main regions of development can be identified, related, respectively, to the two cases defined in Equation (\ref{eqn:pcp_base_flow}). 
In particular, when $\xi R_v\leq \Xi_2$, the effect of wall motion is evidently minor, because while the maximum energy growth generally increases, it does so only to remain around the same order as $G_{\max}$ for crossflow-laden Poiseuille flow (that is, without wall motion). The most noticeable change occurs in a short range around $R_v = R_v^{\mathrm{max}}$, which is highlighted in the inset, although the details here are ultimately immaterial. Despite the substantial increase in $Re_c$ afforded by small $R_v$ coupled with modest wall speeds (as seen, for example, in Figure \ref{fig:critical_parameters}), deemed by \cite{GUHA_2010} as the most relevant parameters for control purposes, non-modal mechanisms nevertheless remain significant and are sometimes even amplified. Of course, the situation can only worsen as $Re$ increases.

\begin{figure}[t!]
\centering
\includegraphics[width=\textwidth]{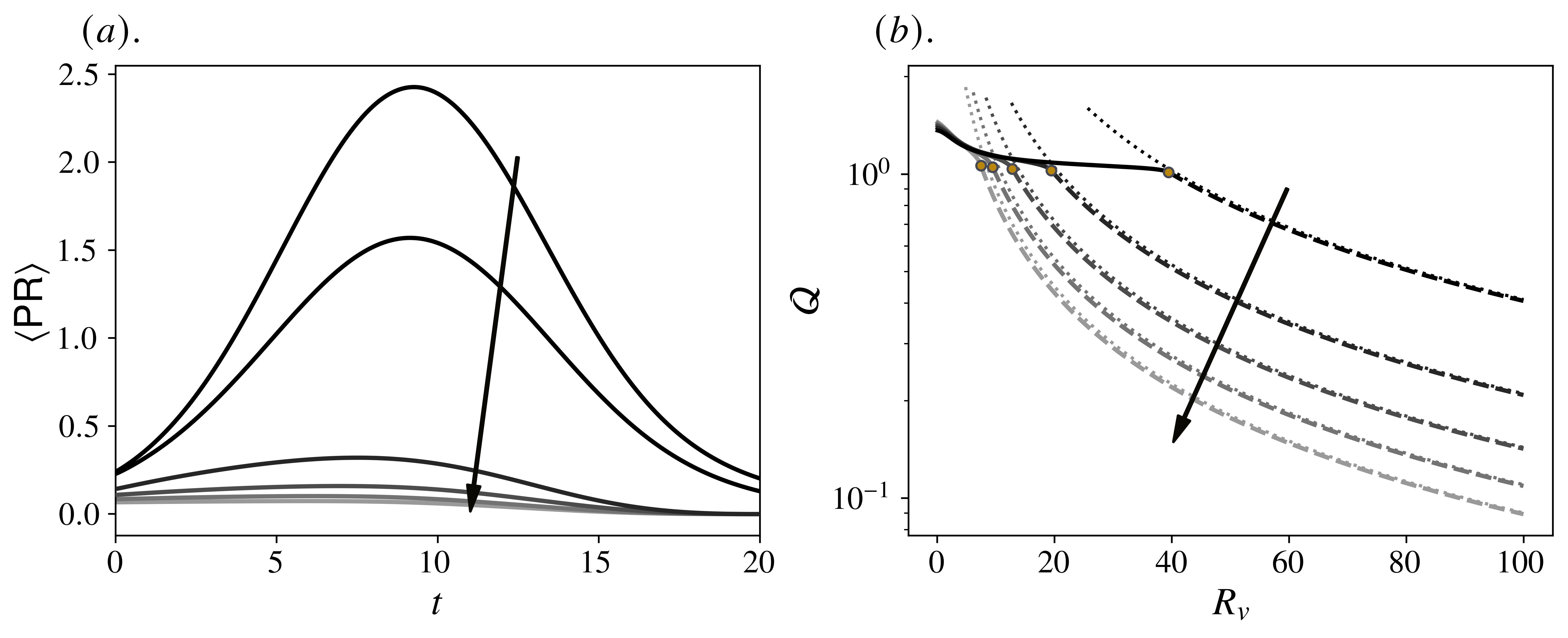}
\caption{$(a$) For the initial condition associated with $G_{\max}$, the evolution of the energy production in time at $R_v = 45$ for the $\xi$ treated in Figure \ref{fig:pcp_gmax}$(a)$ (note that $\xi R_v > 4$ is always satisfied for this choice of the crossflow). $(b)$ The flow rate $\mathcal{Q}$; for each wall speed, the value $R_v^{\mathrm{shift}}$ is marked with a circle, whereas the dashed lines correspond to the regime $\xi R_v > \Xi_2$. The asymptotic result derived in Equation (\ref{eqn:asymp_Q}) is also illustrated (dotted lines). In each panel, $\xi$ increases in the direction of the arrow.}
\label{fig:prods_pcp}
\end{figure} 

Theoretically, one could then look towards flow parameters such that $\xi R_v > \Xi_2$ (Figure \ref{fig:pcp_gmax}$(b)$ plots $R_v = R_v^{\mathrm{shift}}$ initiating this regime) and, indeed, Figure \ref{fig:pcp_gmax}$(a)$ indicates that for sufficiently large wall speeds, successively weaker crossflows can considerably suppress algebraic growth. At first glance, this is encouraging, but it is not without its own caveats. Specifically, an increase in $R_v$ (or $\xi$) for this regime is equivalent to a decrease in the background shear $\mathsf{D}U$, which asymptotically behaves as $\mathsf{D}U\sim  2/\left(\xi R_v\right)$. Consequently, as we highlight in Figure \ref{fig:prods_pcp}$(a)$, the energy production must also decrease. However, what is alarming is that, in this limit, the flow itself is, in fact, being killed off. To verify this, it is easiest to consider the non-dimensional volumetric flow rate $\mathcal{Q}$ which, for $\xi R_v > \Xi_2$, behaves as
\begin{equation}
\label{eqn:asymp_Q}
 \mathcal{Q}\sim\dfrac{1}{R_v} + \dfrac{4}{\xi R_v}
\end{equation}
so that counter-productively, $\mathcal{Q} \to 0$ at large $R_v$; see Figure \ref{fig:prods_pcp} $(b)$.

\section{\label{sec:conclusion} Discussion \& Conclusion}

We have performed a detailed re-assessment of stability in channel flows with crossflow. Previous work has attempted to determine parameter regimes supposedly ideal for delaying transition, and our primary contribution in this work has been to show that non-modal growth may preclude this apparent suitability.

The main focus of our study has been the crossflow-laden Poiseuille flow. Beginning with an eigenvalue (modal) analysis, we provide a global perspective on modal instability by tracking the trajectory of the neutral stability curves in the $\left(\alpha, Re\right)$-space. We show that beyond a discontinuity in the critical parameters, two individual neutral curves begin to co-exist, with distinct properties exemplified through consideration of the corresponding linear energetics. The movement of the critical layers is shown to be highly relevant in shaping the development of this budget. 

From a non-modal perspective, which forms the crux of this paper, the resolvent is first inspected for real frequencies, and later the more general complex case through the $\epsilon$-pseudospectra. Substantial amplification is recovered even at relatively mild sub-critical $Re$, and is only damped for large, possibly unfeasible, crossflows. Similar patterns are observed when treating unforced algebraic growth, with non-modal interactions lasting longer (relative to the reference Poiseuille flow) and being more dangerous at weak $R_v$ -- touted in the previous literature as the ``ideal'' stability configuration. The precise mechanism driving (and suppressing) energy growth is explored by considering a component-wise energy budget. The additional stability provided by large $R_v$ is shown to be due to a combination of decreased energy production and a more active velocity-pressure gradient term, which forces inter-component redistribution of energy in a way that significantly dampens the wall-normal fluctuations and, therefore, the lift-up effect.

This analysis has been extended to account for wall motion, cf. the flow of \cite{GUHA_2010}. Here, we present novel instabilities for a region of parameter space previously thought to be unconditionally linearly stable. These instabilities occur at very high $Re$ but are physically genuine and of the short-wavelength type. Optimal growth is found to only worsen with the inclusion of wall motion, except beyond a regime change in the parameter space defined by $R_v = R_v^{\mathrm{shift}}\left(\xi\right)$, where the background shear decays as $1/(\xi R_v)$ in the channel bulk. Here, the decrease in energy production is nonetheless counter-productive, since it is shown to be accompanied by an impractical cessation of mass transport brought on by a skewing of the velocity profile due to the combined effect of crossflow and wall motion. Collectively, our results suggest that crossflow-based stabilization schemes, at least in internal flows, are unlikely to be effective.

\begin{acknowledgments}
M.A. wishes to thank Professor Jeffrey Oishi for helpful comments that improved the clarity of the material presented here, as well as for providing access to high-performance computing time on the Bates College Leavitt Cluster.
\end{acknowledgments}

\appendix

\section{\label{sec:appendix} Effect of the Mean Wall-Normal Velocity}

\begin{figure}[t!]
\includegraphics[width=\textwidth]{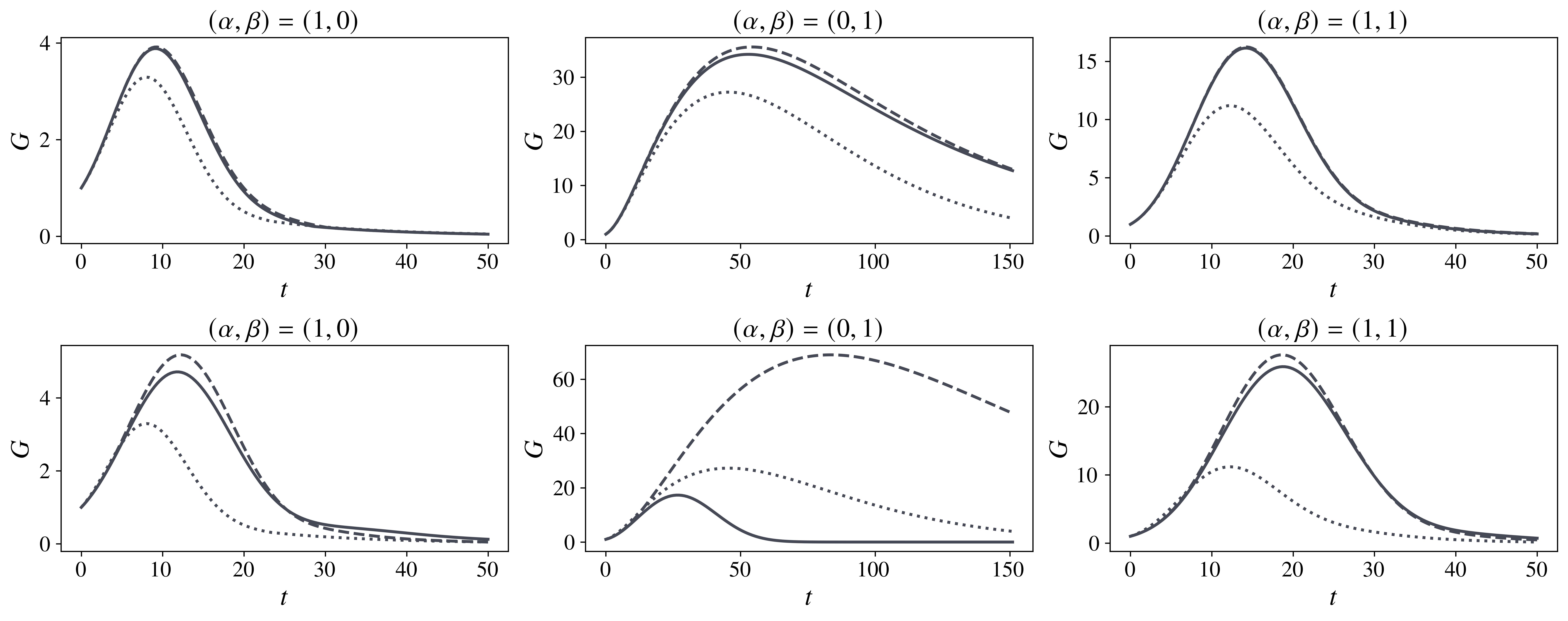}
\caption{The gain $G$ plotted against time for $R_v = 1$ (top) and $R_v = 15$ (bottom) for (left to right), $\left(\alpha, \beta\right) = \left(1, 0\right)$, $\left(\alpha, \beta\right) = \left(0, 1\right)$, and $\left(\alpha, \beta\right) = \left(1, 1\right)$. Solid lines indicate calculations with the crossflow term and dashed lines without. The dotted line in each panel represents Poiseuille flow.}
\label{fig:inertial_operator_differences}
\end{figure}

An intriguing assertion put forward by \cite{FRANSSON2003259} for the asymptotic suction boundary layer was that the reduction in $G$ relative to the reference Blasius case was primarily a consequence of variations in the streamwise velocity profile and not due to the auxiliary terms introduced in the Orr-Sommerfeld-Squire system by the non-zero mean wall-normal velocity. In contrast, \cite{GUHA_2010} determined that the influence of these auxiliary terms on the spectra was, in fact, relevant, although only when $R_v$ was large. 
In a similar manner, here, we wish to categorize the impact of crossflow on algebraic growth between its two main contributions, that is, the inertial operator $\mathcal{I}$
\begin{equation}
\label{eqn:inertial_operator}
    \mathcal{I}\equiv V\dfrac{\partial}{\partial y}\nabla^2
\end{equation}
cf. Equations (\ref{eqn:os}) and (\ref{eqn:sq}) and the asymmetry imposed on the streamwise velocity profile $U$. We follow the previous approaches. In particular, for some sample wavenumber pairs, Figure \ref{fig:inertial_operator_differences} plots the gain $G$ as a function of time for $R_v = 1$, $R_v = 15$, and the Poiseuille flow ($R_v = 0$). For $R_v \neq 0$, two sets of stability equations are considered, one with and the other without the operator $\mathcal{I}$, although both retain the $U$ derived in the presence of crossflow. Therefore, the latter case is equivalent to setting $V = 0$ \textit{a posteriori}. Interesting phenomena are observed; specifically, in all cases, including the crossflow term \textit{dampens} the energy gain relative to when it is excluded. However, in line with the findings of \cite{GUHA_2010}, this disparity only becomes significant for $R_v = 15$. In particular, for the latter, it is apparent that the crossflow-laden $U$ by itself can generate a much stronger response for streamwise-independent disturbances compared to not only Poiseuille flow but also the equivalent profile at $R_v = 1$ (which, as suggested in Figure \ref{fig:gmax} should, in fact, be the more ``dangerous'' scenario). This potential for growth is, of course, significantly suppressed when the inertial operator is taken into account.

Thus, we conclude that the crossflow has a dual impact on non-modal stability. At small $R_v$, its influence is felt predominantly through changes in the background streamwise flow (hence the negligible differences observed in Figure \ref{fig:inertial_operator_differences}). However, for large $R_v$, the third-order differential term in Equation (\ref{eqn:inertial_operator}) is of greater relevance, likely impacting energy amplification directly through variations in the perturbations themselves (and, by extension, the energy budget; see the discussion toward the end of Section \ref{sec:nonmodal_analysis}).

\end{document}